\documentclass[12pt,draftclsnofoot,onecolumn]{IEEEtran}

\usepackage[utf8]{inputenc}
\usepackage[T1]{fontenc}
\usepackage{url}
\usepackage{ifthen}
\usepackage{cite}

\usepackage[cmex10]{amsmath}

\usepackage{graphicx}
\usepackage{amsfonts}
\usepackage{amssymb}
\usepackage{mathrsfs}
\usepackage{subfigure}
\usepackage{mdwmath}
\usepackage{mdwtab}
\usepackage{dsfont}
\usepackage{tcolorbox}
\usepackage[ruled,vlined,linesnumbered]{algorithm2e}

\newcommand\smallO{
	\mathchoice
	{{\scriptstyle\mathcal{O}}}
	{{\scriptstyle\mathcal{O}}}
	{{\scriptscriptstyle\mathcal{O}}}
	{\scalebox{.7}{$\scriptscriptstyle\mathcal{O}$}}
}
\newtheorem{remark}{Remark}
\newtheorem{theorem}{Theorem}
\newtheorem{lemma}{Lemma}
\newtheorem{prop}{Proposition}
\newtheorem{defi}{Definition}
\newtheorem{coro}{Corollary}
\newtheorem{condition}{Condition}

\newenvironment{iarray}{\begin{IEEEeqnarray}{rCl}}{\end{IEEEeqnarray}\ignorespacesafterend}

\begin{document}
\title{Closed-Form Whittle's Index-Enabled Random Access for Timely Status Update
}

\author{Jingzhou Sun, Zhiyuan Jiang, Bhaskar Krishnamachari,~\IEEEmembership{Senior Member,~IEEE}, Sheng Zhou, Zhisheng Niu,~\IEEEmembership{Fellow,~IEEE}
	\thanks{
	
    J. Sun, S. Zhou and Z. Niu are with Beijing National Research Center for Information Science and Technology, Tsinghua University, Beijing 100084, China. Emails: \{sunjz18@mails., sheng.zhou@, niuzhs@\}tsinghua.edu.cn.

    Z. Jiang is now with Shanghai Institute for Advanced Communication and Data Science, Shanghai University, Shanghai 200444, China. He did this work when he was with Tsinghua University. Email: zhiyjiang@foxmail.com.

    B. Krishnamachari is with the Ming Hsieh Department of Electrical Engineering, University of Southern California, Los Angeles, CA 90089, USA. Email: bkrishna@usc.edu. The corresponding author is Zhiyuan Jiang.

    Part of the work has been presented at International Teletraffic Congress ITC 30 \cite{jiang18_itc} in Vienna, Austria.
    }
    }
    \maketitle

\begin{abstract}
We consider a star-topology wireless network for status update where a central node collects status data from a large number of distributed machine-type terminals that share a wireless medium. The Age of Information (AoI) minimization scheduling problem is formulated by the restless multi-armed bandit. A widely-proven near-optimal solution, i.e., the Whittle's index, is derived in closed-form and the corresponding indexability is established. The index is then generalized to incorporate stochastic, periodic packet arrivals and unreliable channels. Inspired by the index scheduling policies which achieve near-optimal AoI but require heavy signaling overhead, a contention-based random access scheme, namely Index-Prioritized Random Access (IPRA), is further proposed. Based on IPRA, terminals that are not urgent to update, indicated by their indices, are barred access to the wireless medium, thus improving the access timeliness. A computer-based simulation shows that IPRA's performance is close to the optimal AoI in this setting and outperforms standard random access schemes. Also, for applications with hard AoI deadlines, we provide reliable deadline guarantee analysis. Closed-form achievable AoI stationary distributions under Bernoulli packet arrivals are derived such that AoI deadline with high reliability can be ensured by calculating the maximum number of supportable terminals and allocating system resources proportionally.
\end{abstract}


\section{Introduction}
Previous generations of wireless networks were hardly incentivized to pursue ultra low-latency or high-reliability communications since human perceptions, e.g., hearing and vision, are insensitive to delay under $100$ milliseconds \cite{liu09}. However, in the foreseeable future, machine-type data will become the dominant traffic contents in wireless networks, replacing human-centric data. Unlike human-centric data (e.g., videos and texts), machine-type data (e.g., sensory data and control message) may have stringent delay and reliability requirements to facilitate novel applications such as autonomous driving, drone swarms, and robotic remote controls. Optimizing latency and reliability is thus of unprecedented interest in 5G and beyond wireless network.

Considerable efforts have been put into optimizing the \emph{end-to-end (E2E) delay} in wireless networks, i.e., delay of a packet through the network between transmission at a source node and reception at a destination node \cite{hou09,kou18}. However, one can argue that this optimization paradigm has two main drawbacks, especially for the upcoming machine-type data dominated era. Firstly, E2E delay does not account for characteristics of information sources and hence treats packets equally. In many scenarios, data packets generated by machine-type information sources have distinct properties and differential priorities, e.g., status information generated by sensors usually exhibits Markovian property, namely a new status renders an old status useless, or a more up-to-date price of a stock is much more valuable than an old one. These characteristics should be judiciously exploited to improve the network efficiency; otherwise, as is indeed encountered today \cite{xy14}, the network wastes precious resources delivering low-value, repetitive packets. The other important aspect is that, as pointed out by several pioneer works \cite{kaul12,kaul11}, E2E delay is a metric from the packet perspective; in practice, we are more concerned with the network level performance, e.g., a low-utilization network wherein individual packets experience a low E2E delay might suffer from insufficient information exchange. 

In view of this, Age of Information (AoI) has been proposed \cite{kaul12,huang15,costa16,najm17,sun17} to specifically characterize \emph{information-to-information (I2I)} remote tracking delay from destination nodes to source nodes. Such an I2I delay, i.e., AoI, can be formally defined to be sufficient to characterize Markovian information sources, e.g., status data sources which constitute a major part of IoT traffic, in the sense that the remote information tracking performance can be uniquely determined by AoI \cite{yin18}. Formally, AoI denotes the difference between the current time and the generation time of current status (maintained at the destination node), i.e., 
\begin{equation}
    \underline{\textrm{AoI at time }t} \triangleq t - \mu(t),
\end{equation} 
where $\mu(t)$ denotes the generation time of the currently newest received status at destination. It is observed from the definition that AoI is concerned from the information destination perspective and can effectively distinguish packet importance---a fresher packet benefits the AoI more. Therefore, AoI is more suitable, especially for Markovian information sources, than E2E delay to be concerned in numerous important IoT applications, e.g., a central controller which requires time-sensitive status parameters from sensors, a cellular base station which utilizes channel state information for efficient transmissions in fast-fading scenarios, and distributed actuators where each actuator requires independent and timely input.

\begin{figure}[!t]
\centering
\includegraphics[width=0.5\textwidth]{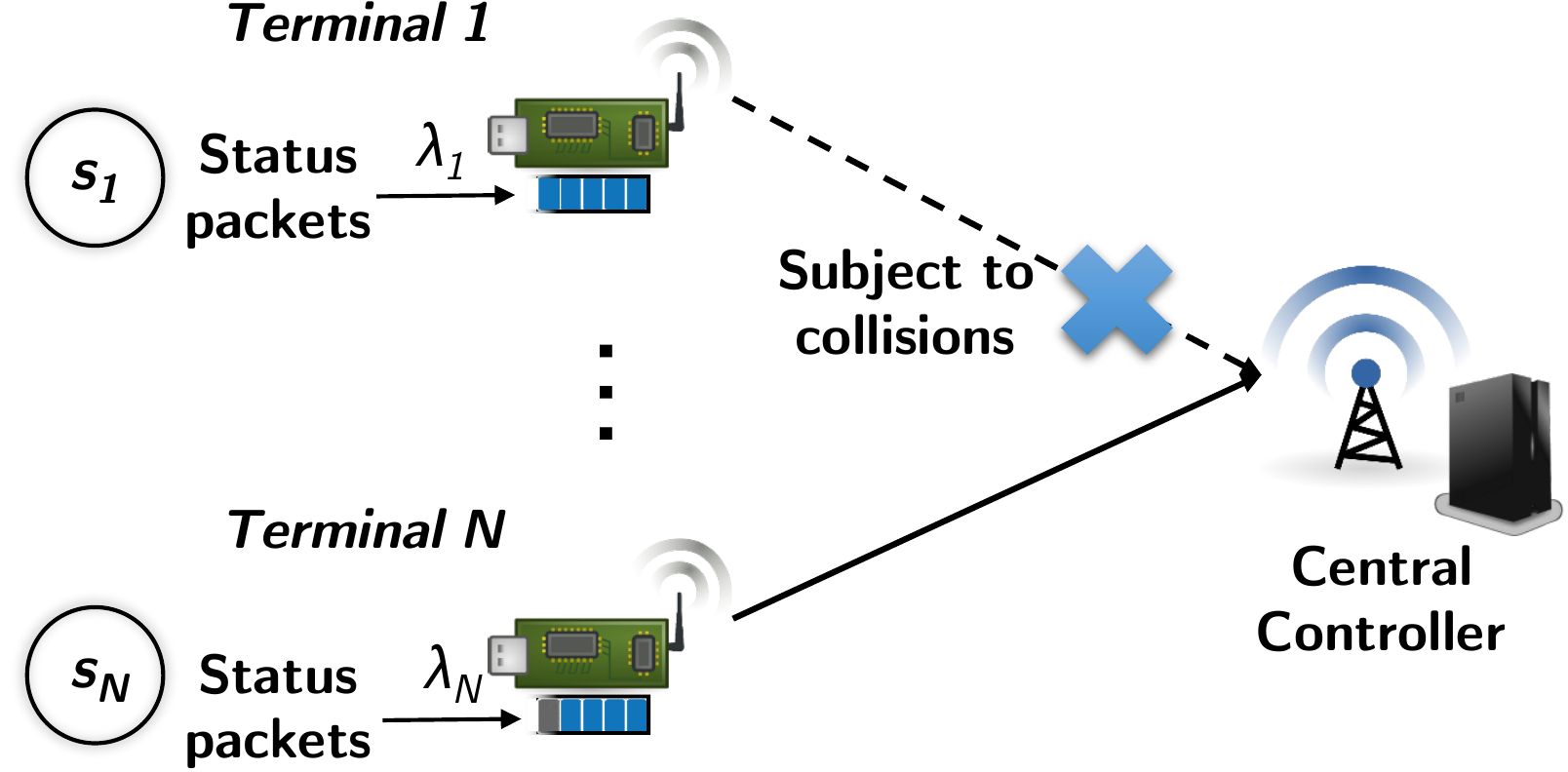}
\caption{A general status update system with $N$ source terminals sharing the wireless uplink.}
\label{fig_arch}
\end{figure}

In this paper, we consider the wireless uplink scenario wherein a large number of terminals, namely massive connectivity, share the wireless medium\footnote{Note that the proposed algorithms can be readily adapted to systems with orthogonal resources, i.e., multiple terminals can transmit simultaneously leveraging e.g., spatial multiplexing, orthogonal subcarriers, by selecting the $k$-highest indices in the Whittle's index policy.} to report status packets to a central controller (see Fig. \ref{fig_arch}). System scalability with massive connectivity \cite{bock18} is thus a big issue---not only is the latency compromised by a large number of terminals sharing the medium, but the signaling overhead necessary for scheduling decisions also constitutes a major concern considering short packets. Therefore, we are particularly interested in contention-based random access schemes, i.e., terminals transmit status update packets autonomously (subject to collisions) without a centralized scheduling process. The main contributions of the paper can be summarized to answer the following three questions:

    1) \emph{How to schedule terminals for optimum time-average AoI?} In this part, our goal is to find the optimal centralized (assuming global information and scheduled access) status update policy which minimizes the long-time average AoI. Towards this end, we derive the Whittle's indices \cite{whitt84} in closed-form, for the scenario with stochastic packet arrivals and reliable channels, and establish the corresponding indexability. This result is further generalized to the cases with unreliable channels and periodic packet arrivals. The Whittle's index policy is near-optimal, based on \cite{web90} and numerous evaluation results in the literature. 
    
    2) \emph{Can decentralized random access achieve the centralized optimum?} A contention-based random access scheme for status update, i.e., Index-Prioritized Random Access (IPRA), is proposed based on the derived Whittle's indices. IPRA prioritizes terminals that are urgent to update based on their locally-computed Whittle's indices in a random access framework. It is shown to be very close to the optimal AoI through computer-based simulation results.
    
    3) \emph{What AoI deadline can be guaranteed with a given reliability requirement?} We further derive closed-form achievable AoI stationary distribution. Based on the result, given AoI deadlines and deadline violation probabilities, the maximum number of supportable terminals can be derived and resources can be allocated accordingly. The results can be instructive for resource management where stringent AoI deadlines, instead of time-average AoI, are concerned.

The rest of the paper is organized as follows. In Section \ref{sec_sm}, we describe the system model and status update design problem in detail. In Section \ref{sec_whittle}, we derive the centralized index policy; specifically, we first formulate the scheduling problem by Restless Multi-Armed Bandit (RMAB) and then derive the Whittle's index and its indexability. Generalizations to periodic packet arrivals and unreliable channels are presented. IPRA is also described for contention-based random access implementation. In Section \ref{sec_dead}, we present an achievable reliable AoI deadline analysis. Section \ref{sec_nr} presents computer-based simulation results. Finally, Section \ref{sec_con} concludes the work.

\subsection{Related Work}
The concept of AoI is related to several existing latency metrics in the literature, e.g., E2E delay \cite{hou09}, inter-delivery-time \cite{singh15}, whereas with notable differences and specific focus on Markov sources as mentioned before. There have been extensive efforts for AoI optimization in various scenarios. Considering a single source-destination pair, the queuing theory has been applied by many works to analyze and optimize the AoI performance under various arrival distributions, service distributions, and queuing policies (cf. \cite{kaul12,najm17,soysal2018age,huang15}); the problem of scheduling multiple sources and multiple flows has been mainly formulated and addressed with Markov Decision Process (MDP) and RMAB under different channel models and system assumptions \cite{kadota16,hsu18,jiang18_iot,bedewy2018age}. There also has some work on AoI in multihop network recently \cite{buyukates2018age,multihop1}. The scaling law of age with respect to the number of nodes in a large network with a hierarchical structure was studied in \cite{buyukates2019age}. Considerable efforts have also been dedicated to considering energy harvesting sources \cite{ara17,arafa2019using} and etc. 

The scheduling problem for AoI optimization in wireless multiaccess networks was considered by Jiang \emph{et al.} \cite{jiang18_iot}, where a round-robin policy with one-packet buffers policy (RR-ONE) was shown to achieve asymptotically optimal performance. However, in the non-asymptotic regime, RR-ONE exhibits notable performance degradation whereas our proposed IPRA is still close to the optimal. Yates and Kaul \cite{yates17} derived closed-form time-average AoI with periodic packet arrivals and unreliable channels, for both round-robin scheduled access and slotted ALOHA random access. However, neither access scheme was optimal and the work did not account for stochastic packet arrivals.

Whittle's index has been adopted to solve the \emph{centralized} scheduling problem for AoI optimization \cite{kadota16,hsu18}. Kadota \emph{et al.} \cite{kadota16} addressed the deterministic packet arrival scenario comprehensively where they derived the index policy and also proved its performance bound. We generalize it to account for stochastic packet arrivals considering the fact that in most scenarios the status packets are generated randomly according to the information source variations, as well as random access implementation. Hsu \cite{hsu18} derived the Whittle's index with random packet arrivals, however, a specific no-buffer packet management policy was considered whereby all undelivered packets are discarded leading to performance loss. Our work generalizes the work \cite{hsu18} to one-packet buffer case, which is also optimal among all packet buffering policies, and shows evident performance gain by keeping the latest packet instead of discarding it. Moreover, the index derivation in this paper is more challenging due to the fact that a two-dimensional system state is involved (the states in \cite{hsu18} are considered almost one-dimensional since the queuing delay of a packet is either zero or one). Additionally, it is shown that the index expression in \cite[Theorem 7]{hsu18} coincides with a special case of our results. 

The Medium Access Control (MAC) protocols have been studied extensively for IoT systems with massive connectivity \cite{aloha}. Among those, many works have been dedicated to enhance the Carrier-Sense Multiple Access with Collision Avoidance (CSMA/CA) scheme for contention-based random access \cite{laya14,wang15}. Our proposed IPRA behaves similarly with the access barring scheme \cite{wang15}, which rejects terminal access randomly when the number of terminals is large and has been identified as an important enhancement for current random access channel with massive connectivity \cite{3gpp11}. However, IPRA---instead of barring access randomly (or based on a predefined priority class)---can measure the terminal update urgency by its Whittle's index and reject access thereby.

As far as we know, this is the first work to solve the Whittle's index with two-dimensional system states, and the first to achieve near-optimal performance based on a lightweight contention-based random access design.
\begin{table}[!t]
	\caption{Description of Key Notations}
	\centering
	\begin{tabular}{ l  l}
		\hline
		$N$: & Number of terminals. \\		
		\hline
		$n$: & Terminal index. \\
		\hline
		$\lambda_n$: & Status packet arrival rate of terminal-$n$. \\
		\hline
		$\boldsymbol{\pi}$: & An admissible policy.  \\
		\hline		
		$\omega_n$: & AoI weight of terminal-$n$.\\
		\hline
		$h_{n,\boldsymbol{\pi}}(t)$ or $h$: & AoI of terminal-$n$ at time $t$ under policy $\boldsymbol{\pi}$.\\	
		\hline
		$u_{n,\boldsymbol{\pi}}(t)$ or $u$ & Scheduling decision for terminal-$n$ under policy $\boldsymbol{\pi}$ at time $t$. \\
		\hline
		$a_{n,\boldsymbol{\pi}}(t)$ or $a$: & Queuing delay of the packet in terminal-$n$'s buffer at time $t$ (one-packet buffer).\\
		\hline
		$d_{n,\boldsymbol{\pi}}(t)$ or $d$: & Difference between AoI of terminal-$n$ $(h_{n,\boldsymbol{\pi}}(t))$ and $a_{n,\boldsymbol{\pi}}(t)$.\\
		\hline
		$\hat{J}^*$: & Optimal average cost of the decoupled problem.\\
		\hline
		$m$: & Auxiliary service charge in the decoupled problem. \\
		\hline
		$f(a,d)$: & Differential cost function with state $(a,d)$ in the decoupled problem.\\
		\hline
		$I_{\mathsf{b},n}(\cdot)$: & Whittle's index of terminal-$n$ with Bernoulli packet arrivals and reliable channels.\\
		\hline
		$I_{\mathsf{p},n}(\cdot)$: & Whittle's index of terminal-$n$ with periodic packet arrivals and reliable channels.\\
		\hline
		$I_{\mathsf{e},n}(\cdot)$: & Whittle's index of terminal-$n$ with unreliable channels.\\
		\hline
		$p_{\mathsf{e},n}$: & Transmission failure probability of terminal-$n$.\\
		\hline
		$\epsilon_n$: & Deadline violation probability of terminal-$n$.\\
		\hline
		$H_n$: & AoI Deadline of terminal-$n$.\\
		\hline
	\end{tabular}
	\label{tab_notation}
\end{table}

\section{System Model and Problem Formulation}
\label{sec_sm}
We consider a scenario where one central controller is collecting status update packets from $N$ terminals. Key denotations are listed in Table \ref{tab_notation}. A time-slotted system is considered. The status updates are conveyed by randomly generated packets at each terminal, reflecting the current status information sensed by terminals and stored at terminal queues. The packet arrivals are modeled by independently identically distributed (i.i.d.) Bernoulli processes (generalized to periodic arrivals in Section \ref{sec_gene}) with mean rates $\lambda_n\in [0,1]$. Concretely, the $T$-horizon time-average AoI of the system is defined by
\begin{equation}
\label{AoI}
    \Delta_{\boldsymbol{\pi}}^{(T)} \triangleq \frac{1}{T N}\sum_{t=1}^T \sum_{n=1}^N \omega_n \mathbb{E}[h_{n,\boldsymbol{\pi}}(t)],
\end{equation}
where $\boldsymbol{\pi}$ denotes an admissible policy, $\omega_n$ is a pre-defined weight representing the importance of terminal-$n$, $T$ is the time horizon length, and $h_{n,\boldsymbol{\pi}}(t)$ denotes the AoI of terminal-$n$ at the $t$-th time slot under policy $\boldsymbol{\pi}$. The long-time average AoI and the AoI deadline reliability constraint are defined by
\begin{equation}
\label{aoi_inf}
    \bar{\Delta}_{\boldsymbol{\pi}} \triangleq \limsup_{T \to \infty} \Delta_{\boldsymbol{\pi}}^{(T)},
\end{equation}
\begin{equation}
    \limsup_{T \to \infty}\mathbb{E} \left[\frac{\sum_{t=1}^T \mathds{1}(h_{n,\boldsymbol{\pi}}(t)>H_n)}{T} \right]\le \epsilon_n,
\end{equation}
respectively, where $H_n$ is the AoI deadline of terminal $n$, $\epsilon_n$ is the deadline violation time ratio threshold, and $\mathds{1}(\cdot)$ is the indicator function.
\subsection{Status Update Process}
In order to minimize the weighted-average AoI, the terminals should decide the transmission order by which they can update the status in a timely fashion and also avoid collision. Considering only scheduled updates, the status update decisions include:

1) \emph{Scheduled terminal}: Decide which terminal to transmit a status update packet.

2) \emph{Packet management}: Once a terminal is scheduled, a status update packet is then transmitted based on packet management---the terminal can choose one packet from its packet buffer to transmit.

For contention-based random access, the status update process is facilitated by a contention period, which is illustrated in Section \ref{sec_ipra}.

Due to the Markovian property of information sources, i.e., new packet brings in smaller AoI, it is obvious that the optimal packet management policy is to transmit the most up-to-date packet; this is equivalent to maintaining a \emph{one-packet buffer} at each terminal and only retaining the newest packet. Since there is only one buffer, we adopt scalar $a_{n,\boldsymbol{\pi}}(t)$ to represent the queuing delay of the buffer packet in terminal-$n$ at time slot $t$. Note that the one-buffer packet management is not necessarily optimal when considering service interruption \cite{najm16} which, however, does not exist in our system setting.

The evolution of the AoI of terminal-$n$ can be written as
\begin{iarray}
\label{evo}
&& h_{n,\boldsymbol{\pi}}(t+1) =  h_{n,\boldsymbol{\pi}}(t) + 1 - u_{n,\boldsymbol{\pi}}(t) \prod_{j \neq n} (1-u_{j,\boldsymbol{\pi}}(t)) d_{n,\boldsymbol{\pi}}(t),
\end{iarray}
where $u_{n,\boldsymbol{\pi}}(t)=1$ when the terminal-$n$ transmits in this time slot and zero otherwise, and the AoI reduction is denoted by $d_{n,\boldsymbol{\pi}}(t)$ which equals the time duration (time slots) between the generation of the last received packet from terminal-$n$ and the updated packet's generation time, i.e., $d_{n,\boldsymbol{\pi}}(t) = h_{n,\boldsymbol{\pi}}(t) - a_{n,\boldsymbol{\pi}}(t)$. Note that $d_{n,\boldsymbol{\pi}}(t)=0$ if terminal-$n$ has no packet to update. With a slight abuse of notation, we prescribe $a_{n,\boldsymbol{\pi}}(t) = h_{n,\boldsymbol{\pi}}(t)$ in this circumstance until a new packet arrives at the terminal.

To be clear, we elaborate on the sequence of events in one time slot as below.
\begin{figure}[!h]
\centering
\includegraphics[width=0.7\textwidth]{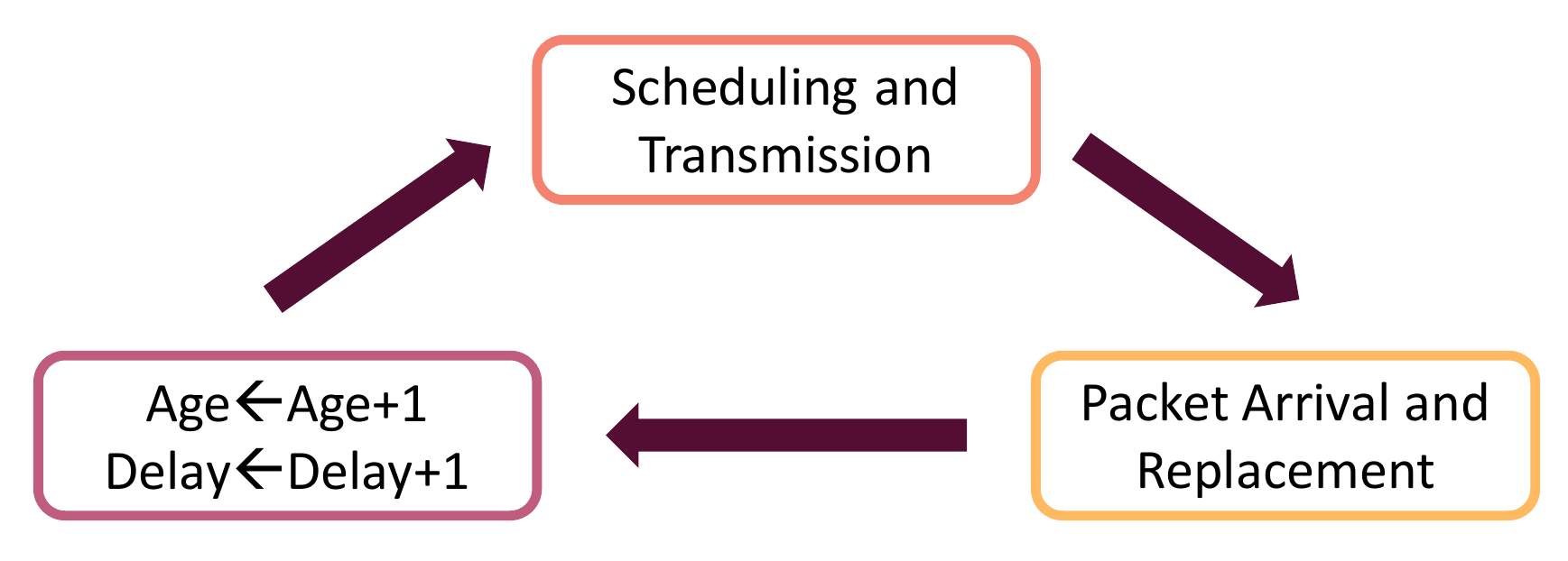}
\caption{The sequence of events in one time slot.}
\label{fig_sq}
\end{figure}
At the beginning of the time slot, AoI $h_{n,\boldsymbol{\pi}}(t)$ and queuing delay $a_{n,\boldsymbol{\pi}}(t)$ increase by 1. Then, we make scheduling decision and transmit a packet. Packet arrival happens after the transmission. If a new packet arrives at terminal-$n$, it will take on the buffer and replace the undelivered old packet. 

\section{Whittle's Index Policy and IPRA}
\label{sec_whittle}
In this section, we first formulate the general status update scheduling problem into an MDP problem which is solved by relative value iteration to give a performance benchmark in the simulation section. Afterward, we solve the decoupled model, where each terminal is examined separately, and thereby derive the Whittle's index. Inspired by the index policy, a contention-based random access scheme with little signaling overhead and comparable performance as the centralized policy is proposed.
\subsection{MDP-Based Problem Formulation}
The system state is denoted by $\mathcal{S}(t) \triangleq \{(a_1(t),d_1(t)),\cdots,(a_N(t),d_N(t))\}$, where $d_n(t)$ is defined as
\begin{equation}
    d_n(t) \triangleq h_{n}(t) - a_n(t).
\end{equation}
Note that this state definition is equivalent to the one with $(a_n(t),h_n(t))$, whereas much more convenient in the following derivations and therefore adopted. Based on the fact that $h_{n}(t) \ge a_n(t)$, it is clear that $d_n(t) \ge 0$, $\forall n,t$; in addition, the queuing delay of buffered packet $a_n(t) \ge 1$, $\forall n,t$, since we make scheduling decision after the increment stage according to Fig.\ref{fig_sq}. The action space is $\mathcal{U} =  \{1,\cdots,N\}$, which denotes the scheduled terminal index. The state transition for terminal-$n$, when not scheduled, is described by
\begin{iarray}
&& \Pr\{(a_n,d_n) \to (a_n+1,d_n) \} = 1-\lambda_n; \nonumber\\
&& \Pr\{(a_n,d_n) \to (1,d_n + a_n) \} = \lambda_n,
\end{iarray}
and when scheduled,
\begin{iarray}
&& \Pr\{(a_n,d_n) \to (a_n+1,0) \} = 1-\lambda_n; \nonumber\\
&& \Pr\{(a_n,d_n) \to (1,a_n) \} = \lambda_n,
\end{iarray}
with $a_n \ge 1$ and $d_n \ge 0$. The objective in this section is to minimize the long-time average AoI:
\begin{equation}
\label{mini}
    \min_{\boldsymbol{\pi} \in \mathcal{U}^T} \limsup_{T \to \infty} \frac{1}{T N}\sum_{t=1}^T \sum_{n=1}^N \mathbb{E}\left[a_{n,\boldsymbol{\pi}}(t) + d_{n,\boldsymbol{\pi}}(t)(1-u_{n,\boldsymbol{\pi}}(t))\right].
\end{equation}
Such an MDP problem can be solved by the relative value iteration method with average cost function \cite{bertsekas2007dynamic}, however, the curse of dimensionality and lack of insights limit the effectiveness of the solution. To address this issue, we note that the above problem can be essentially viewed as an RMAB problem whereby each arm represents one terminal and the reward of pulling an arm is the AoI reduction of the corresponding terminal. It is well-known that the Whittle's index policy is near-optimal for RMAB problems with a large number of arms\cite{web90}. Therefore, we seek for the index policy in the following subsection. 
\subsection{Decoupled Model and Index Policy}
To design the index policy, following Whittle's methodology, a decoupled model is formulated where the time-varying AoI cost (or negative value) of each terminal is compared with a terminal (arm) with a constant cost $m$. For some states $(a,d)$, pulling either arm can lead to the same long-term cost. Then, for each state $(a,d)$, we can find a value $m(a,d)$, which can be interpreted as the minimum service charge that the system is willing to pay for pulling the arm. Therefore, each arm can be investigated separately, and hence the complexity of finding a solution decreases from exponential with $N$ (MDP value iteration) to linear with $N$. Mathematically, considering the minimization in \eqref{mini} subject to the constraint of $\sum_{n=1}^N u_{n,\boldsymbol{\pi}}(t) =1$, $\forall t=1,2,\cdots$, a relaxed constraint can be expressed as
\begin{equation}
    \limsup_{T \to \infty}{\frac{1}{T} \mathbb{E}\left[\sum_{t=1}^T \sum_{n=1}^N u_{n,\boldsymbol{\pi}}(t) \right]} \le 1,
\end{equation}
i.e., the constraint of only scheduling one terminal at a time is relaxed to a time-average constraint. It has been shown that the Whittle's index policy solves the relaxed problem exactly \cite{whitt84} and is near-optimal for the original problem in general. 

However, the main challenge is that the Whittle's index is only defined for problems that are indexable, meaning that the value of each state of an arm can be fully characterized by the constant service charge of the index policy. The existence of such indexability is problem-dependent and usually difficult to establish, especially with multi-dimensional system states whereby simple structures of the solution (e.g., threshold-based) may not exist. We manage to overcome the difficulty and derive the closed-form Whittle's index as follows.

Since only one terminal is considered, we omit the terminal index in this subsection. Concretely, the decoupled problem is formulated by adding a constant service charge $m$ whenever the terminal is scheduled, the objective function of the decoupled model is therefore
\begin{equation}
    \hat{J}^* = \min_{u(t) \in \{0,1\}} \frac{1}{T}\sum_{t=1}^T \mathbb{E}[a(t) + d(t)(1-u(t)) + m u(t)].
\end{equation}
We consider the long-time average where $T \to \infty$. The action is binary, i.e., $u(t)=0$ or $u(t)=1$. This MDP is an average cost problem with infinite horizon and countably infinite state space, and hence the existence of a deterministic and stationary optimal policy is problem-dependent. However, we note that this specific problem can be shown to have an optimal deterministic and stationary policy by checking the conditions in \cite[Proposition 4.6.1]{bertsekas2007dynamic}. For this end, we solve the Bellman's equation and derive an optimal policy consequently. The Bellman's equation is given by
\begin{iarray}
\label{c2go}
f(a,d) + \hat{J}^*  =  \min \left\{ \begin{array}{l}
d + a + (1 - \lambda )f(a + 1,d) + \lambda f(1,d + a),\\
a + m + (1 - \lambda )f(a + 1,0) + \lambda f(1,a)
\end{array} \right\}, 
\end{iarray}
where the upper term in the minimization corresponds to idle and the lower denotes scheduled. The optimal average cost is denoted by $\hat{J}^*$, and $f(a,d)$ is the differential cost-to-go function and we prescribe $f(1,0)=0$. In what follows, we will solve the above Bellman's equation directly. 

\begin{theorem}
\label{thm1}
Considering the decoupled model, given an auxiliary service charge $m$, there exists a stationary policy $\boldsymbol{\pi}_{\mathsf{D}}$ that is optimal over all policies. Under policy $\boldsymbol{\pi}_{\mathsf{D}}$, the action with a state $(a,d)$ is to schedule the terminal when $d \ge D_a$ and idle otherwise. Specifically, the thresholds are given by
\begin{iarray}
\label{da_th}
     D_a = \left\{\,
        \begin{IEEEeqnarraybox}[][c]{l?s}
        \IEEEstrut	
        \left\lceil (1-\lambda+a\lambda)\hat{J}^* - a + 1 - \lambda \frac{a(a-1)}{2} -\frac{1}{\lambda} \right\rceil, &  if $1 \le a < D_1$;\\
    	\left\lceil \lambda m \right\rceil, & if $a \ge D_1$,
    	\IEEEstrut
    	\end{IEEEeqnarraybox}
    	\right.  
\end{iarray}
where $D_1 = \left\lceil \hat{J}^* -\frac{1}{\lambda} \right\rceil$. The optimum average AoI $\hat{J}^*$ is the unique positive solution to the following equation:
\begin{equation}
    m = \left(D_1-1+\frac{1}{\lambda}\right) \hat{J}^* -\frac{D_1^2}{2} + \frac{D_1}{2} -\frac{D_1}{\lambda} + \frac{\lambda-1}{\lambda^2}.
\end{equation}
\end{theorem}
\begin{IEEEproof}
The basic methodology is to first assume that a stationary policy $\pi_D$ with the aforementioned threshold structure that is optimal. With this assumption, we obtain a scalar $\hat{J}^*$ and a function $f$ that satisfy Bellman's equation \eqref{c2go} and prove that $\pi_D$ attains the minimum in \eqref{c2go} for all state. By verifying the condition in \cite[Proposition 4.6.1]{bertsekas2007dynamic}, the optimality of $\pi_D$ is established. See Appendix \ref{app_thm1} for details.
\end{IEEEproof}
\begin{figure}[!t]
\centering
\includegraphics[width=0.6\textwidth]{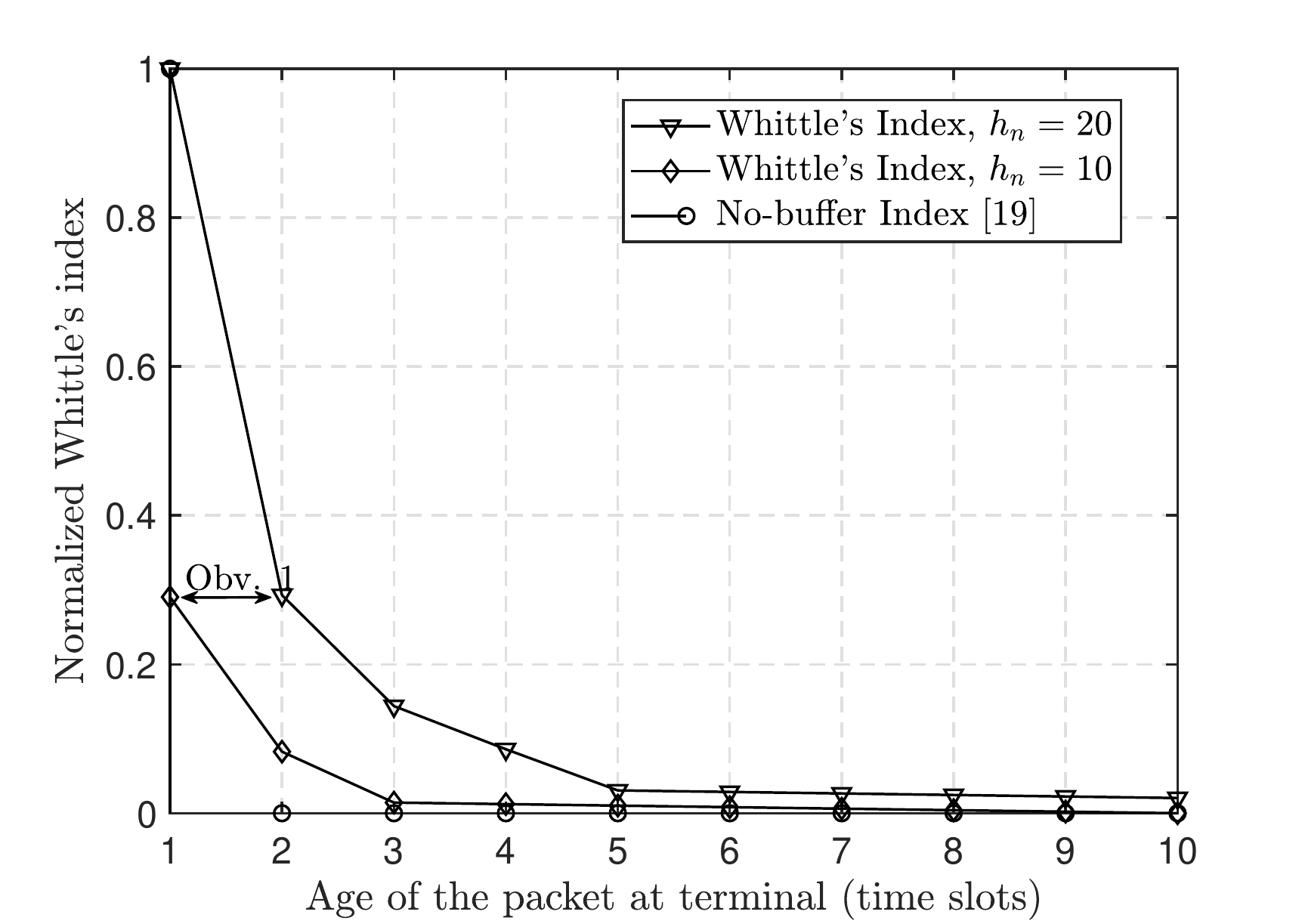}
\caption{Illustration of the difference between the Whittle's indices with optimal buffering strategy and no-buffer.}
\label{fig_indC}
\end{figure}
\begin{remark}
Examining the difference between Theorem \ref{thm1} and \cite[Theorem 5]{hsu18}, due to the no-buffer assumption in \cite{hsu18}, the system states are nearly one-dimensional (queuing delay is either one or no packet) and hence there is only one AoI threshold; in contrast, there is one threshold for \emph{each} $a$ (age of the packet at terminal-side) in Theorem \ref{thm1}, making the derivation of the theorem considerably more challenging. On the other hand, we will show based on simulation results that by using only one-packet buffers, the performance gain is significant compared with the no-buffer index policy; the reason can be specified by comparing the normalized indices with various $a$; note that only the relative value of the index is relevant in making scheduling decision hence justifying the normalization. The benefit of buffering is shown by the fact that the index value of a state with the packet's queuing delay $a$ larger than one is still significant; hence dropping the packets of such kind compromises performance. In particular, based on observation $1$ in Fig. \ref{fig_indC}, a state with $(a,h)=(1,10)$ has approximately the same index with a state $(2,20)$, meaning that scheduling a terminal of state $(2,20)$ (buffering the packet for one time slot) is equally beneficial as one of $(1,10)$, whereas the no-buffer index deems the state $(2,20)$ as worthless. $\hfill\square$
\end{remark}

The indexability of the index policy can be readily derived based on Theorem \ref{thm1}.
\begin{defi}[Indexability]
Given costs $m_1$ and $m_2$, and the sets of states of which the optimal action is to idle are denoted by $\Pi_{m_1}$ and $\Pi_{m_2}$ respectively, the problem is indexable if 
\begin{equation}
    \forall m_1, m_2\textrm{ with } m_1 < m_2 \Rightarrow \Pi_{m_1} \subseteq  \Pi_{m_2},
\end{equation}
and for $m=0$, $\Pi_{m} =  \emptyset$; for $m \to \infty$, $\Pi_{m}$ is the entire state space.
\end{defi}
\begin{theorem}[Indexability]
\label{thm2}
Consider the decoupled model and the scheduling policy $\boldsymbol{\pi}_{\mathsf{D}}$ given in Theorem \ref{thm1}, then $\boldsymbol{\pi}_{\mathsf{D}}$ is indexable. $\hfill\square$
\end{theorem}
\begin{IEEEproof}
See Appendix \ref{app_thm2} for details.
\end{IEEEproof}

The index for any state is described as follows, based straightforwardly on Theorem \ref{thm1}.
\begin{theorem}
\label{def_index}
Consider the decoupled model and denote the index by $I_\mathsf{b}(a,d,\lambda)$ with state $(a,d)$ and arrival rate $\lambda$, 
\begin{tcolorbox}[title = Whittle's Index with Bernoulli Arrivals and Reliable Channels]
\begin{iarray}
\label{index_e}
I_\mathsf{b}(a,d,\lambda) = \left\{\,
        \begin{IEEEeqnarraybox}[][c]{l?l}
        \IEEEstrut	
        \omega \left(\frac{1}{2} x^2 + \left(\frac{1}{\lambda}-\frac{1}{2}\right) x\right), &\textrm{if } d > \frac{\lambda}{2}a^2+\left(1-\frac{\lambda}{2}\right)a; \\
    	\omega \frac{d}{\lambda},&\textrm{otherwise}, 
    	\IEEEstrut
    	\end{IEEEeqnarraybox}
    	\right. 
\end{iarray}
\end{tcolorbox}
\noindent where $x\triangleq \frac{d + \frac{a(a-1)}{2}\lambda}{1-\lambda+a\lambda}$ and $\omega$ is terminal weight. $\hfill\square$
\end{theorem}
\begin{remark}
The derivation of the index follows from the reasoning that the index of a state equals the minimum auxiliary service charge that makes the scheduling decisions of the terminal under the current state equally appealing. Later in Section \ref{sec_gene}, it will be generalized to periodic packet arrivals and unreliable channels. $\hfill\square$
\end{remark}
\begin{remark}
The index is a generalization of previous results in \cite{kadota16} where the index without randomly generated status packets is derived. In their work, the index is (with transmission success probability $p=1$, user weight $\alpha=1$ and frame length $T=1$ in \cite{kadota16})
\begin{iarray}
\label{h1}
I(h) = \frac{1}{2}h(h+1),
\end{iarray}
where $h$ is the AoI of the terminal. Based on \eqref{index_e}, with $\lambda=1$ and hence the packet age is $a=1$, we obtain
\begin{iarray}
\label{h2}
I_\mathsf{b}(1,d) &=& \frac{1}{2}d^2 + \left(\frac{1}{\lambda}-\frac{1}{2}\right)d = \frac{1}{2} d(d+1).
\end{iarray}
The difference between $h$ in \eqref{h1} and $d=h-1$ in \eqref{h2} is due to the fact that \cite{kadota16} adopts the pre-action age and we adopt the post-action age; either case does not affect the results much. The derived index also coincides with \cite{hsu18} when $a=1$. Therefore, it is observed that our result is consistent with previous works and generalize to the scenario with random packet arrivals and one-buffer strategy. $\hfill\square$
\end{remark}
\subsection{Generalizations to Periodic Packet Arrivals and Unreliable Channels}
\label{sec_gene}
In this subsection, Whittle's index for periodic packet arrivals---instead of Bernoulli arrivals assumed before---and for unreliable channels are derived respectively, based on similar methodologies adopted in Theorem \ref{def_index}. In many real-world applications, the status updates are generated regularly, e.g., sensors generate sampled sensory data in a periodic manner and send them. In this regard, the Whittle's index is derived as follows.
\begin{coro}
\label{coro_T}
Denote the index with periodic packet arrivals at terminals by $I_\mathsf{p}(a,n_\mathsf{p},T_\mathsf{p})$ where the packet arrival interval is $T_\mathsf{p}$ and the state $(a,n_\mathsf{p})$ denotes that the queuing delay of the packet at terminal is $a$ where $1 \le a \le T_\mathsf{p}$ because packets periodically arrive with interval $T_\mathsf{p}$, and $(n_\mathsf{p} T_\mathsf{p}+a)$ time slots have passed since the last update where $n_\mathsf{p} \in \{0,1,2,\cdots\}$. 
\begin{tcolorbox}[fonttitle=, title=Whittle's Index with Periodic Arrivals and Reliable Channels]
\begin{iarray}
\label{index_t}
I_\mathsf{p}(a,n_\mathsf{p},T_\mathsf{p}) = T_\mathsf{p}^2 (\left\lfloor K_1 \right\rfloor +1)\left(K_1-\left\lfloor\frac{K_1}{2}\right\rfloor\right),
\end{iarray}
\end{tcolorbox}
\noindent where $K_1 \triangleq  \frac{n_\mathsf{p}(T_\mathsf{p}-a+1)}{T_\mathsf{p}}$. $\hfill\square$
\end{coro}
\begin{IEEEproof}
The index expression with periodic arrivals can be viewed as a simplified version of Theorem \ref{def_index} where the arrivals are stochastic, due to the fact that the arrival process is deterministic. The proof is also based on Theorem \ref{def_index} with necessary changes.
\end{IEEEproof}
\begin{remark}
Note that this result differs from Kadota \emph{et al.} \cite{kadota16} which ignores the AoI increment inside a transmission frame of length $T_{\mathsf{p}}$. With the Whittle's indices derived in Theorem \ref{def_index} and Corollary \ref{coro_T}, we can optimize the AoI of a network consisting of nodes with heterogeneous packet arrival patterns, namely deterministic and stochastic, by scheduling them based on indices derived respectively. This is due to the decoupling of terminals in the Whittle's index approach. $\hfill\square$
\end{remark}
\begin{coro}
\label{coro_e}
Denote the index with unreliable channels by $I_{\mathsf{e},n}$ and the transmission failure probability by $p_{\mathsf{e},n}$. $I_{\mathsf{e},n}$ can be approximated by the index without transmission error, denoted by $I_{\mathsf{reliable},n}$, multiplied by $1-p_{\mathsf{e},n}$, and the approximation error tends to zeros as $p_{\mathsf{e},n} \to 0$, i.e.,
\begin{tcolorbox}[title = Whittle's Index with Unreliable Channels]
\begin{equation}
\label{index_p}
    I_{\mathsf{e},n} \overset{p_{\mathsf{e},n} \to 0}{\longrightarrow}  (1-p_{\mathsf{e},n})I_{\mathsf{reliable},n} + \smallO(1),
\end{equation}
\end{tcolorbox}
\noindent where $I_{\mathsf{reliable},n}$ can be specified by, e.g., Theorem \ref{def_index} or Corollary \ref{coro_T}.
\end{coro}
\begin{IEEEproof}
This asymptotic result has an insightful structure: A node with transmission error probability $p_{\mathsf{e},n}$ (i.i.d.) takes approximately $1/(1-p_{\mathsf{e},n})$ consecutive time slots to reach a successful transmission; a successful transmission is worth $I_{\mathsf{reliable},n}$, in terms of service charge the system is willing to pay in the index formulation, and hence $I_{\mathsf{e},n}$ is equal to $I_{\mathsf{reliable},n}$ divided by the average trial-and-error times $1/(1-p_{\mathsf{e},n})$. A sketch of the proof is given in Appendix \ref{app_coro_e}.
\end{IEEEproof}
\subsection{Index-Prioritized Random Access}
\label{sec_ipra}
The index policy derived above clearly requires global information (all the age and AoI information) for scheduling decisions, and hence it is recognized as a centralized policy with scheduled access which makes it undesirable for short-packet status update in massive IoT systems due to signaling overhead concerns. Nonetheless, inspired by the index policy, we describe IPRA whose performance is close to that of the centralized index policy, however, with lightweight contention-based random access protocol structure. 

\begin{figure}[!t]
\centering
\includegraphics[width=0.7\textwidth]{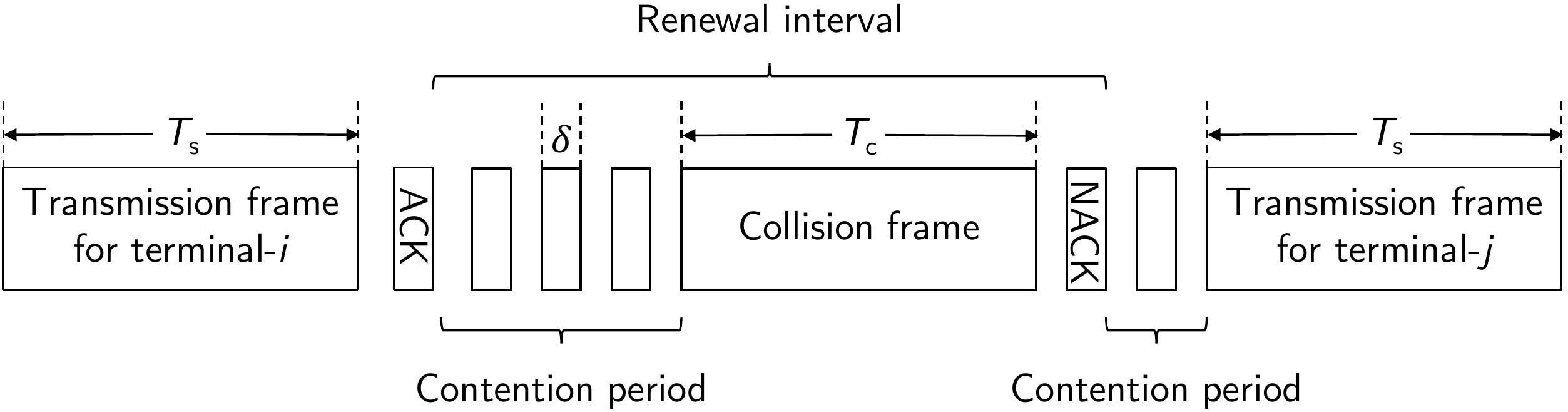}
\caption{Transmission structure of IPRA.}
\label{fig_ipra}
\end{figure}

IPRA works roughly as follows. Given its transmission history and packet arrivals, each terminal can calculate its index $I_n$ based on \eqref{index_e}, \eqref{index_t} or \eqref{index_p}, depending on the channel and arrival property. Thereby, this individual index is mapped to a transmission probability based on a public mapping function which captures the idea that only valuable packets (packets with high index values) are transmitted; a random access (contention) period is hence introduced to resolve possible collisions. The selection of the public mapping function is tricky and we propose to use a single-threshold function which, notwithstanding its simplicity, achieves near-optimal performance based on simulation results. The detailed procedure is described in the box below, as well as the frame structure in Fig. \ref{fig_ipra}. The single-threshold structure of IPRA makes it behave similarly with the Access Class Barring (ACB) enhancement for Random Access Channel (RACH) in the 3GPP community; ACB also rejects access to a subset of terminals, i.e., zero transmission probability in IPRA. However, ACB rejects terminals randomly, irrespective of their transmission urgency, whereas IPRA selects terminals that are the most urgent to update, thanks to the derived Whittle's index.

\begin{tcolorbox}[title=Algorithm 1: IPRA]
\textbf{{Contention period: For $n \in \{1,\cdots,N\}$}}\\
\If{$I_n \ge \mathsf{indexThreshold}$}{Terminal-$n$ transmits with probability $p$.}
\Else{Terminal-$n$ is idle.}
\textbf{Transmission/Collision frame:}\\
If the transmission is successful, the central controller feeds back an ACK; otherwise a NACK is fed back.\\
Go to the contention period.	
\end{tcolorbox}

In IPRA, two parameters, i.e., transmission probability $p$ and index threshold $\mathsf{indexThreshold}$, need to be optimized. In our implementation, we find that the optimization is usually easy since the performance, as a function of these parameters i.e. $f(p,\mathsf{indexThreshold})$, is quasi-concave (unimodal). Therefore, well-known methods such as \cite{comb14} can be applied. In fact, in our experiments, simply fixing $p$ and a one-dimensional search on $\mathsf{indexThreshold}$ renders satisfactory performance.

Furthermore, the overhead of random access of IPRA is asymptotically negligible when the packet transmission time is long compared with one time slot, based on a similar argument as in \cite{gai11} which shows that the overhead of standard CSMA/CA is negligible by adjusting the access probability $p$---for IPRA, $p$ is optimized by a search algorithm. In reality, the transmission of a packet usually takes multiple time slots (a packet transmission time of $100$ is common in current Wi-Fi), and hence the consideration of this asymptotic regime is justified. Note that although the packet is relatively long in this regime, the overhead of centralized scheduling is still significant since it is related to the number of terminals. Denote the packet transmission time and collision frame length as $T_\mathsf{s}$ and $T_\mathsf{c}$ respectively, and $\delta$ as the length of one contention time slot (herein $\delta$ equals one time slot). We assume $T_\mathsf{s} = T_\mathsf{c}$ which is the worst-case assumption meaning that a terminal can only find out the transmission fails after one entire transmission frame, i.e., no collision detection is considered. The performance of IPRA is evaluated in Section \ref{sec_nr}.

\section{Achievable Reliable Deadline Region Analysis}
\label{sec_dead}
In previous sections, we focus on optimizing the long-time average AoI. In some practical use cases such as alert information, there is a hard deadline of AoI, beyond which the status information is much less valuable or even worthless. In this regard, it is of great interest to understand the AoI deadline violation probability and develop algorithms which can ensure reliable AoI deadline accordingly.

As discovered in our previous work \cite{jiang18_isit}, the minimum time-average AoI in wireless uplinks (collision-based interference model), even with centralized scheduling, is proportional to the number of terminals with a scaling factor of $\frac{1}{2}$. Therefore, it is unreasonable to require a low AoI deadline with a large number of terminals. Because of this, understanding the relationship between the deadline violation probability and the number of terminals sharing the wireless medium is important. To this end, we first analyze the stationary AoI distribution of a terminal with deterministic periodic scheduling times, i.e., a terminal that is scheduled every $\Gamma$ time slots. The stationary distribution leads to an algorithm that can arrange the number of terminals sharing one channel to guarantee that the deadlines are met with high probability. 

\begin{theorem}
\label{thm_sta}
Assuming Bernoulli arrivals at a terminal and the scheduling interval is a constant of $\Gamma$ time slots, the steady-state stationary cumulative distribution of the AoI is
\begin{tcolorbox}[title=Achievable Stationary Distribution of AoI]
\begin{equation}
\label{eq_dist}
    F(x) = \left\{\,
    \begin{IEEEeqnarraybox}[][c]{l?s}
    \IEEEstrut	
	\frac{x-\frac{1-\lambda}{\lambda}\left(1-(1-\lambda)^x\right)}{\Gamma}, & $1 \le x \le \Gamma$;\\
	1-(1-\lambda)^{x-\Gamma+1} \frac{1-(1-\lambda)^\Gamma}{\lambda\Gamma}, & $x \ge \Gamma+1$, 
	\IEEEstrut
	\end{IEEEeqnarraybox}
	\right.
\end{equation}
\end{tcolorbox}
\noindent where $F(x) \triangleq \Pr\{h \le x\}$ where $h$ is the steady state AoI.  $\hfill\square$
\end{theorem}
\begin{IEEEproof}
The proof follows from \cite[Theorem 3]{jiang18_iot} and the details are omitted for brevity.
\end{IEEEproof}
\begin{remark}
Note that a necessary condition for nodes to have deterministic scheduling intervals in the collision-based wireless uplinks (assuming reliable channels) is $\sum_{n=1}^{N} \frac{1}{\Gamma_n} \le 1$, where $\Gamma_n$ is the scheduling interval of node $n$.
However, this condition is not sufficient, since the constant scheduling intervals of multiple nodes, as required in the theorem, may not be satisfied due to possible scheduling collisions, e.g., two nodes with $\Gamma_1=2$ and $\Gamma_2=3$ may collide every $6$ time slots. Nonetheless, the influence of this imperfection on the performance is considered minimum after slightly manipulating the scheduling times. $\hfill\square$
\end{remark}
\begin{coro}
\label{coro_dl}
An achievable reliable deadline region can be characterized by
\begin{tcolorbox}[title=Achievable Reliable Deadline Region]
\begin{equation}
    \sum_{n=1}^N \frac{\log(1-\lambda_n)}{W_{-1}\left(\frac{\log(1-\lambda_n)}{\epsilon_n c(\lambda_n,H_n)}\right)} \le 1,
\end{equation}
\end{tcolorbox}
\noindent where $W_{-1}(\cdot)$ is the negative branch of the Lambert $W$ function \cite{lambert}, the AoI deadline is denoted by $H_n$, $c(\lambda_n,H_n) \triangleq \frac{\lambda_n}{(1-\lambda_n)^{H_n+1}}$, and the reliability requirement is $\epsilon_n$, i.e., $\Pr\{h>H_n\} \le \epsilon_n$, which can be made arbitrarily small.

When $N$ is large, and assume $\epsilon_n=\epsilon$, $\lambda_n=\lambda$, $H_n = H$, in order to achieve a time-average AoI of $H$, or an AoI deadline of the same value $H$, the numbers of terminals, i.e., $N_\mathsf{mean}$ and $N_\mathsf{deadline}$, should satisfy respectively:
\begin{iarray}
\label{dl_mean}
N_\mathsf{mean} \le 2 H,\textrm{ and }N_\mathsf{deadline} &\le& H - \frac{\log\epsilon}{\log(1-\lambda)}+ c_0(\lambda),
\end{iarray}
where $c_0(\lambda)\triangleq1+\frac{\log(-\log(1-\lambda))-\log \lambda}{\log(1-\lambda)}$. $\hfill\square$
\end{coro}
\begin{IEEEproof}
See Appendix \ref{app_coro_dl}.
\end{IEEEproof}
\begin{remark}
Considering that terminals can occupy orthogonal system resources and hence be divided into several groups, scheduling with a deadline requires, at least, approximately twice the resources. $\hfill\square$
\end{remark}

Theorem \ref{thm_sta} and Corollary \ref{coro_dl} can be leveraged to determine the maximum number of supportable terminals of the system, given their deadlines and reliability coefficients $\epsilon$. Since these regions are proved achievable based on deterministic scheduling intervals, those deadlines can be guaranteed with high reliability.

\section{Simulation Results}
\label{sec_nr}
\begin{figure}[!t]
    \centering    
    \subfigure[]{\includegraphics[width=0.24\textwidth]{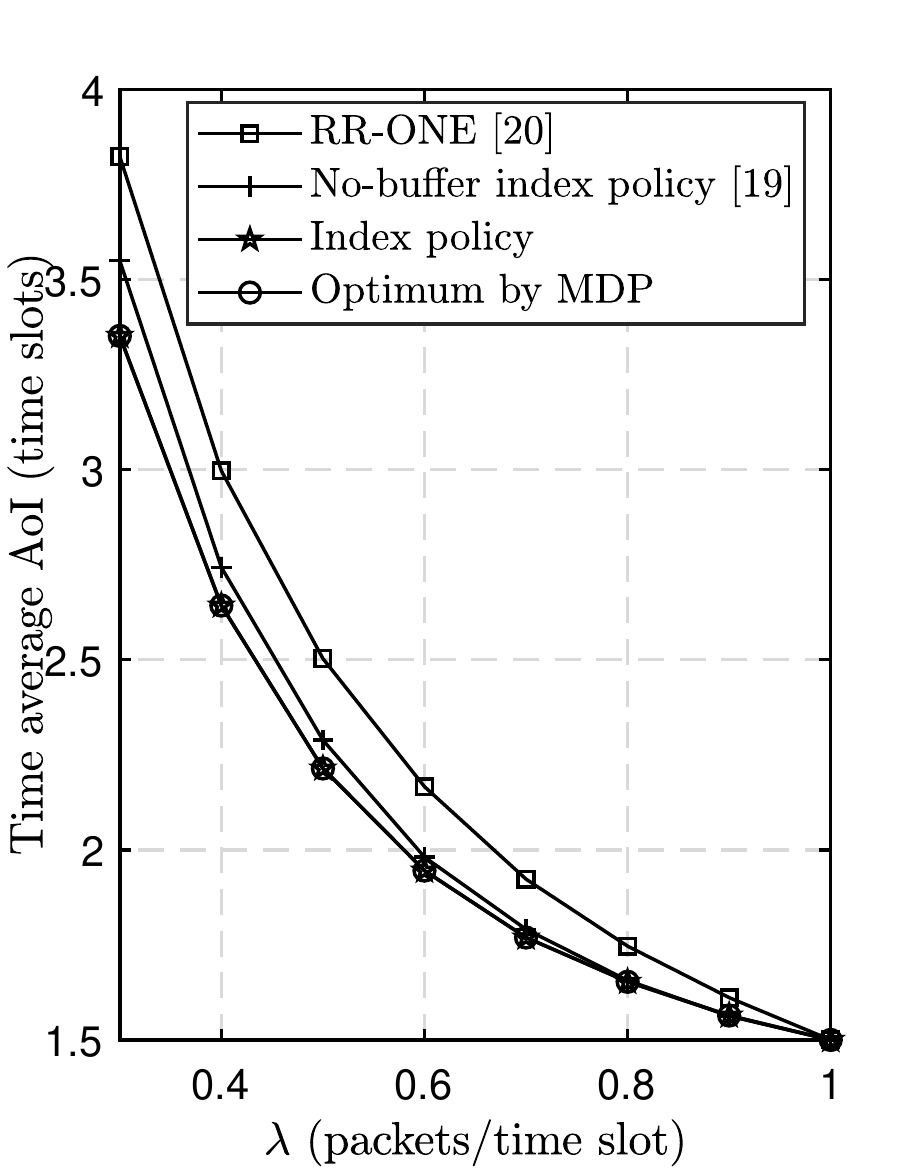}}
    \subfigure[]{\includegraphics[width=0.24\textwidth]{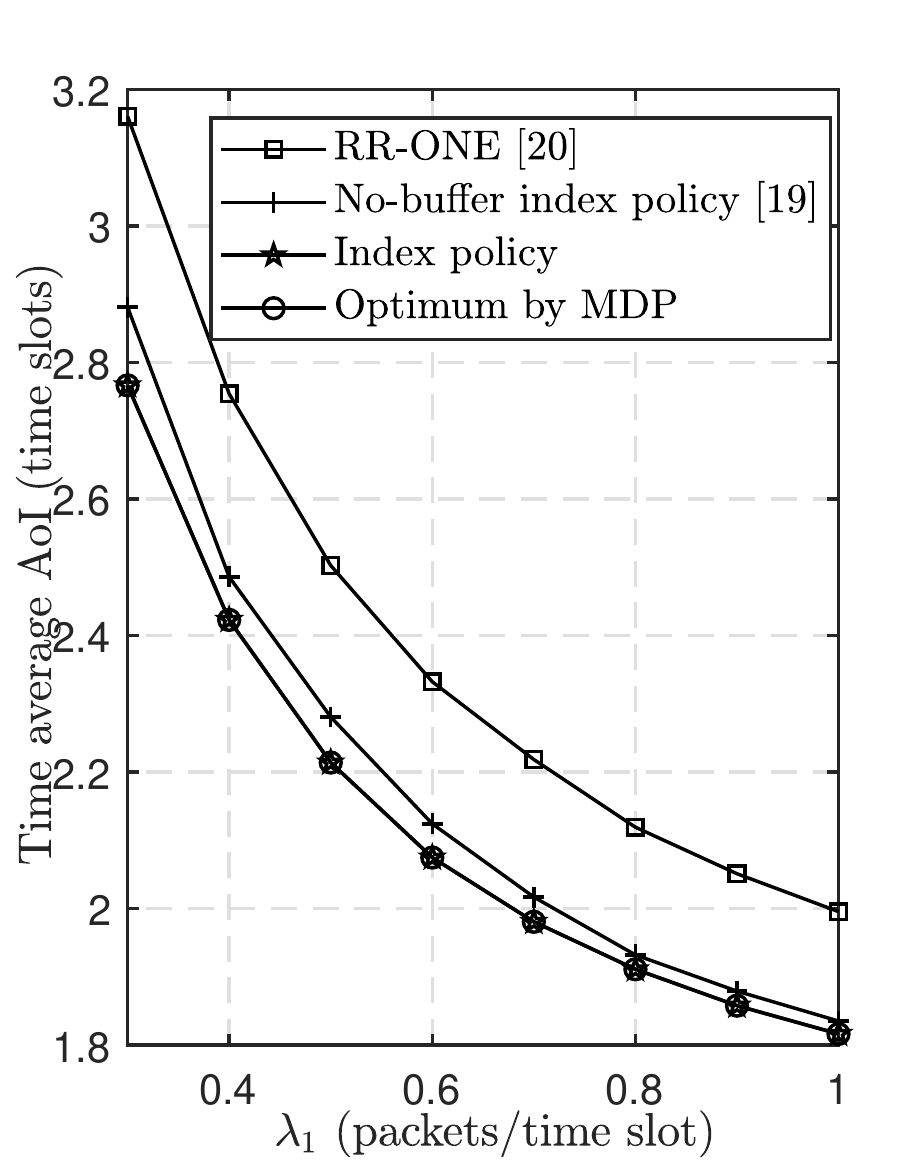}}
    \subfigure[]{\includegraphics[width=0.24\textwidth]{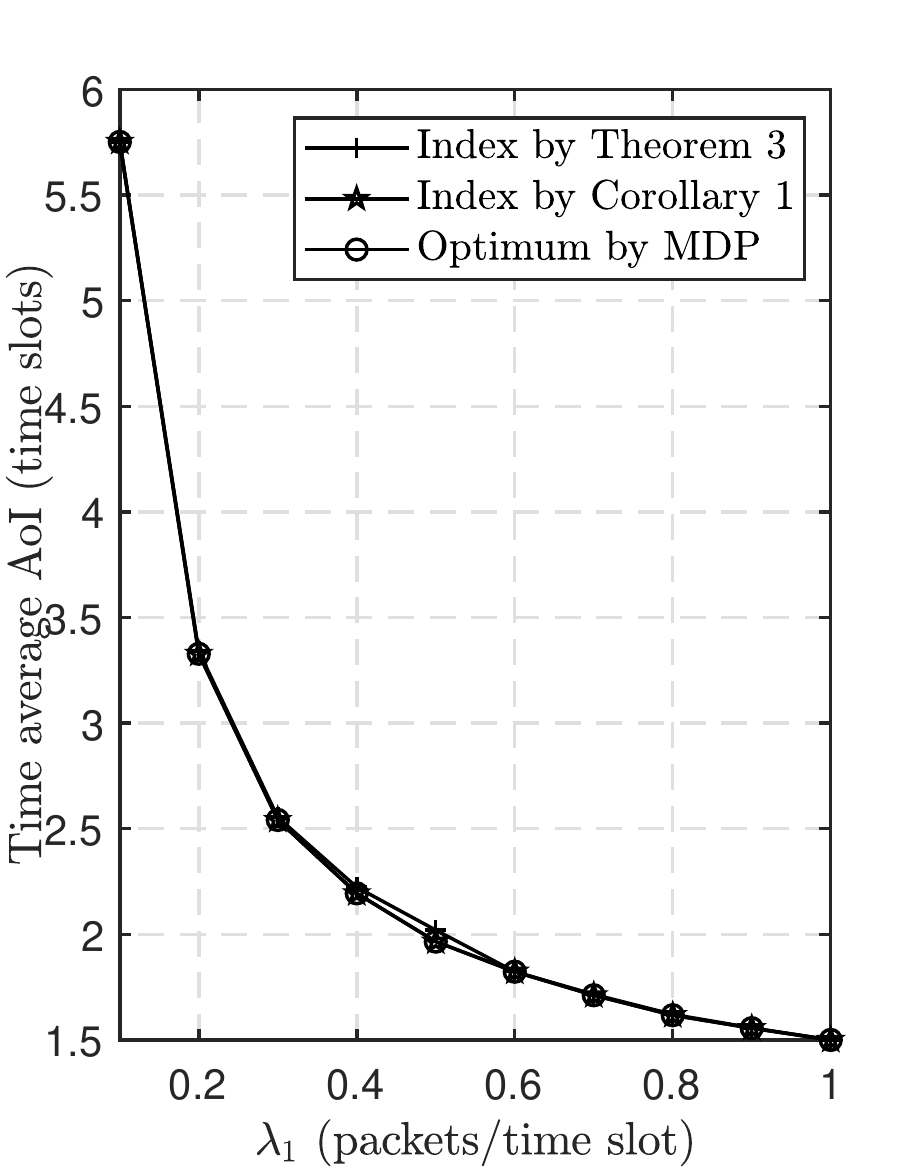}}
    \subfigure[]{\includegraphics[width=0.24\textwidth]{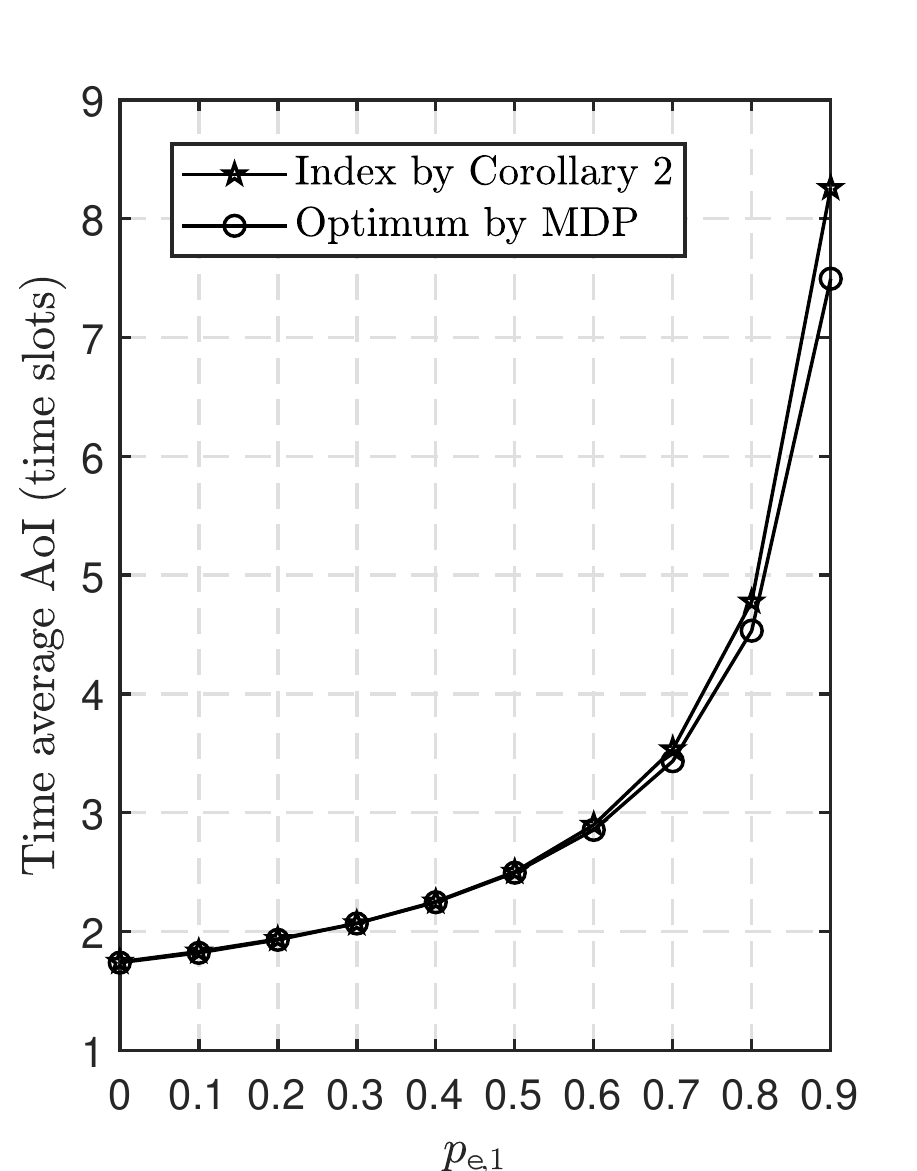}}
    \caption{Performance comparisons with optimum by solving the MDP with $2$ terminals: (a) identical packet arrival rate with reliable channels; (b) heterogeneous arrival rates with terminal $1$'s arrival rate shown as x-axis and terminal $2$'s fixed as $0.5$ (reliable channels); (c) heterogeneous packet arrival patterns with terminal $1$ being stochastic and terminal $2$ periodic $(T_\mathsf{p}=2)$ (reliable channels); (d) unreliable channels with $p_{\mathsf{e},2}=0.9$ and $\lambda_1=\lambda_2=0.8$.}
    \label{fig_mdp}
\end{figure} 
\begin{figure}[!t]
\centering
\includegraphics[width=0.6\textwidth]{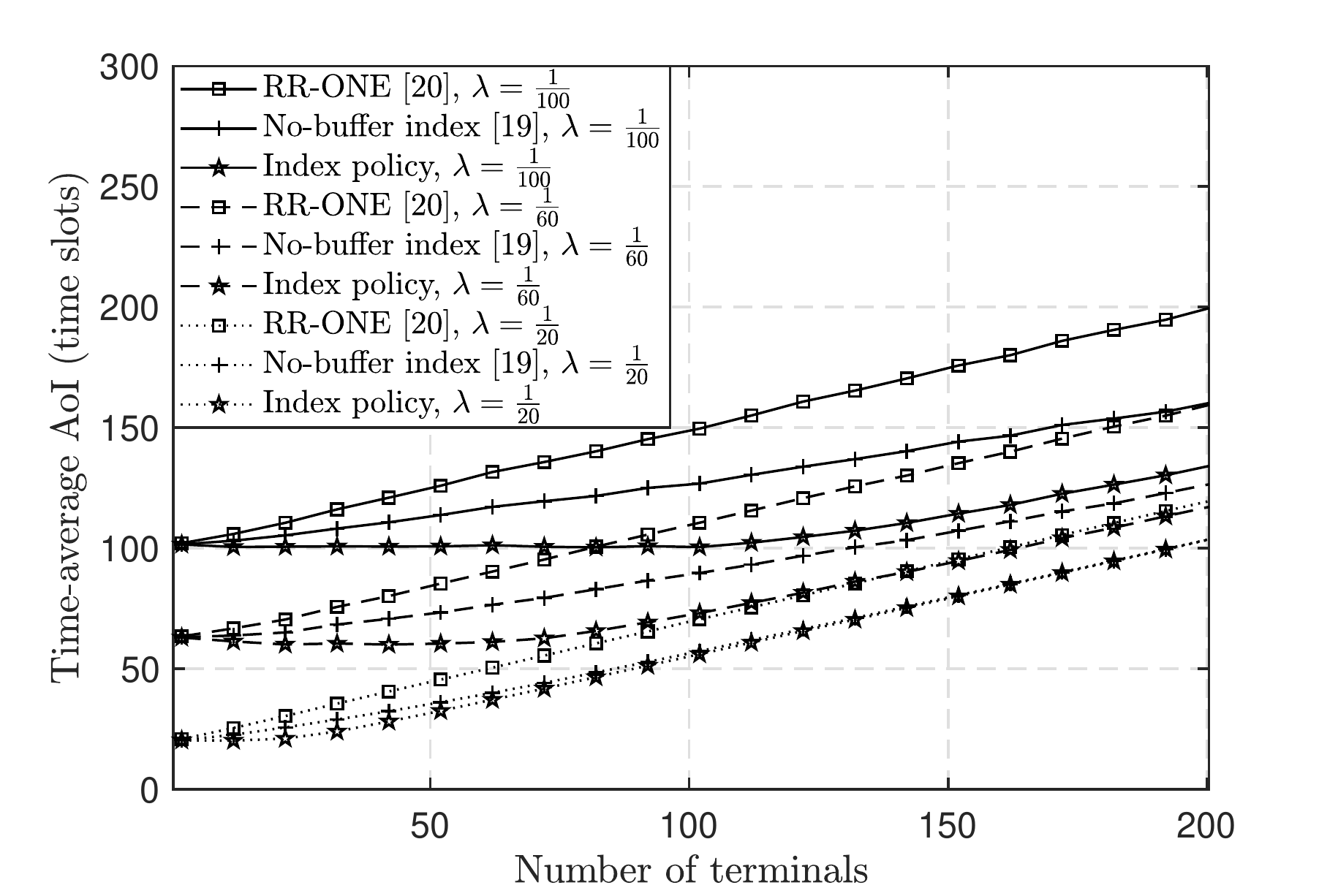}
\caption{Performance comparisons with a large number of terminals and identical packet arrival rate of $\lambda$ specified in the legend.}
\label{fig_indN}
\end{figure}
\begin{figure}[!t]
    \centering    
    \subfigure[]{\includegraphics[width=0.45\textwidth]{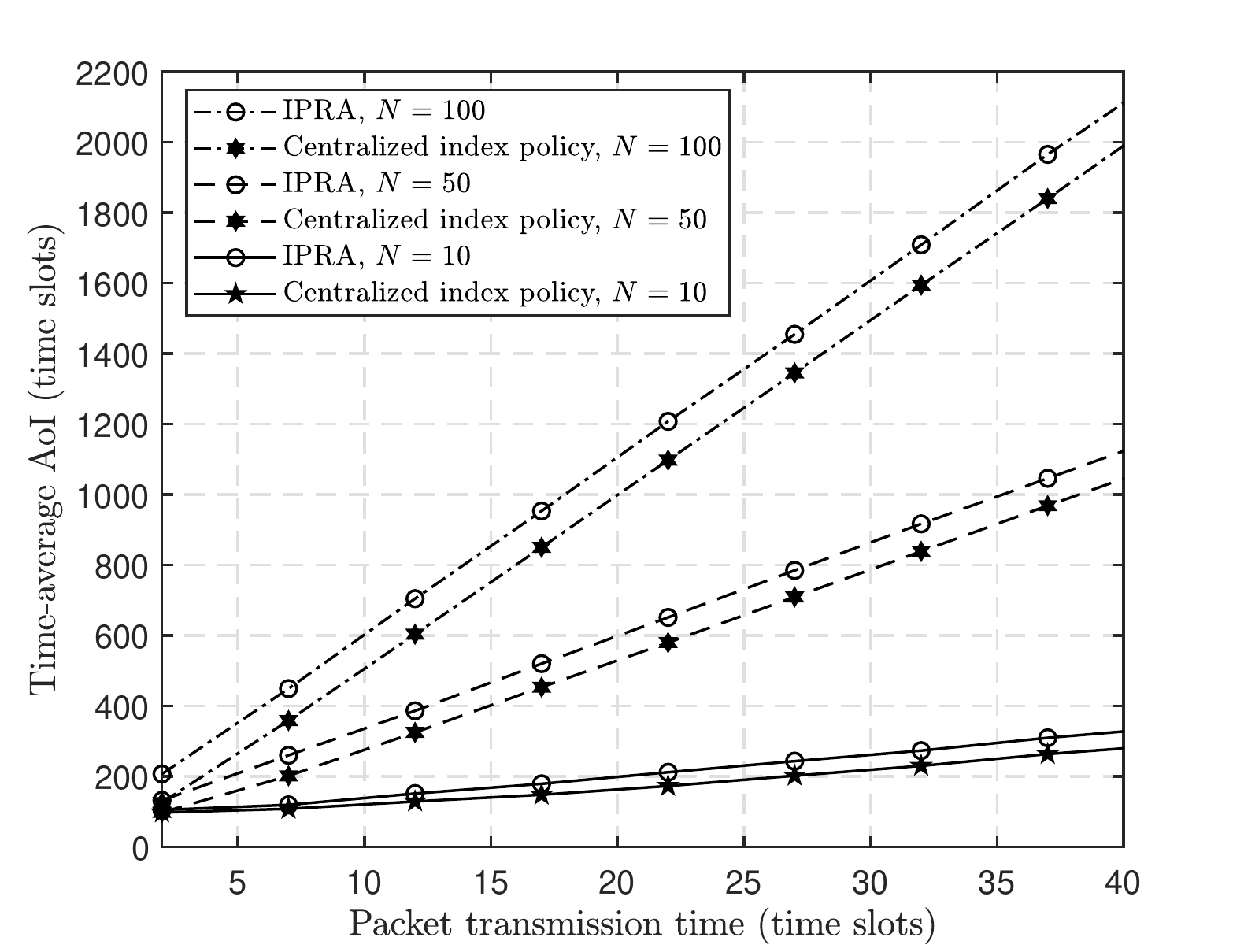}}
    \subfigure[]{\includegraphics[width=0.45\textwidth]{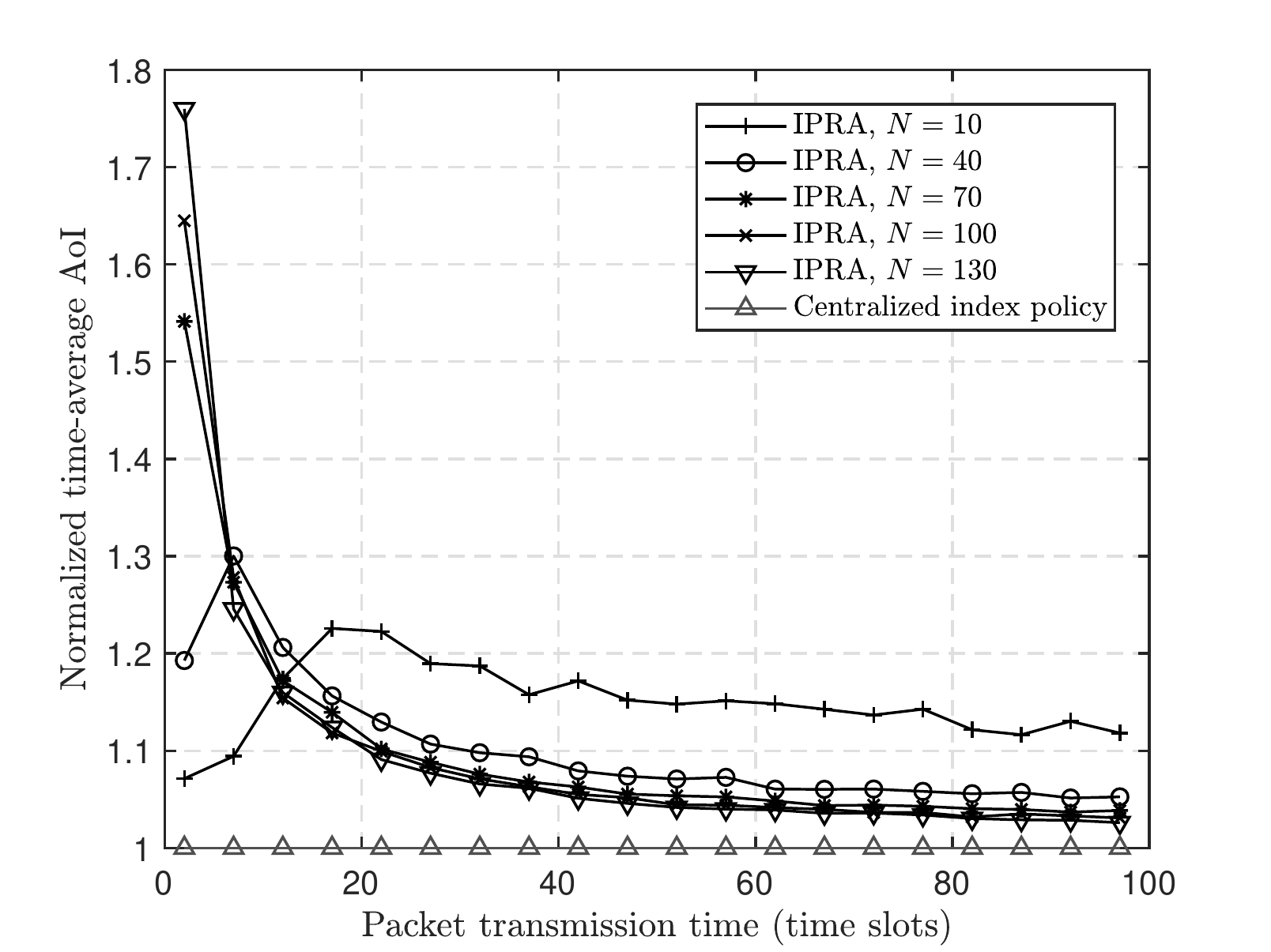}}
    \caption{Performance evaluations for IPRA. The packet arrival rate is $\lambda_n=0.01$, $\forall n \in \{1,\cdots,N\}$ and the channels are reliable.}
    \label{fig_ipra_perf}
\end{figure} 
In this section, we present simulation results which are based on running the scheduling policies for $10^6$ time slots and obtaining the time-average AoI.
In Fig. \ref{fig_mdp}, a $2$-terminal case, due to the curse of dimensionality, is considered and the optimum AoI based on solving the MDP via relative value iteration is obtained. Based on Fig. \ref{fig_mdp} (a) and Fig. \ref{fig_mdp} (b), the difference between the Whittle's index policy and the optimum is hardly visible, which is in line with various existing works \cite{kadota16,hsu18}. However, the no-buffer index policy \cite{hsu18} suffers evident performance loss when the packet arrival rate is low. This can be explained that the no-buffer strategy drops precious (when with low arrival rates) status packets when both terminals have arrivals in the same time slot, in which case packets should be stored for update in the future. Note that the no-buffer index policy is identical with the index policy in this paper when $\lambda$ approaches one, by showing that the index in Theorem \ref{thm2} in this regime, i.e., \eqref{h2}, coincides with the index in \cite{hsu18}.\footnote{The difference between $h$ here and $h+1$ in \cite{hsu18} is due to the difference in the definitions of initial packet age.} In Fig. \ref{fig_mdp} (c), we consider heterogeneous packet arrival patterns, i.e., one terminal has Bernoulli-based stochastic arrivals whose index is calculated based on Theorem \ref{thm2} and the other has periodic packet arrivals with index given by Corollary \ref{coro_T}. It is observed that, again, the index policy achieves close to optimal performance; moreover, the index policy seems insensitive to packet arrival patterns, by observing that using indices given by Theorem \ref{thm2} for both terminals with one terminal having deterministic packet arrivals makes negligible performance degradation. Fig \ref{fig_mdp} (d) investigates how well is the index approximation in Corollary \ref{coro_e} with unreliable channels. The approximation applies to a wide range of channel error probability, as it is shown that only when $p_{\mathsf{e}}>0.7$ can we observe noticeable performance degradation. 

The performance advantage of the proposed index policy (Theorem \ref{thm2}) over existing works is more pronounced in Fig. \ref{fig_indN} with a larger network scale. It was shown that RR-ONE \cite{jiang18_iot} achieves the optimal scaling factor with a large $N$ but is suboptimal with finite $N$; this is observed in the 
. Combining with the fact that the time-average AoI achieved by a standard CSMA/CA protocol with equal transmission probability is approximately twice that of RR-ONE \cite{jiang18_iot}, the index policy outperforms standard CSMA/CA significantly. The gap between no-buffer and the proposed one-buffer policy is the most evident when the mean arrival interval is comparable with the number of terminals, i.e., what we identify as the \emph{joint-asymptotic regime} where $N \to \infty,\,\frac1\lambda \to \infty$, and $N\lambda \to C$, and $C$ is a fixed constant. This is because, in this regime, the delays due to scheduling among terminals and random packet arrivals are equally significant and hence neither can be ignored, making the seek for optimal policy most challenging; on the other hand, when $N \to \infty$ with fixed $\lambda$, the delay due to random packet arrivals can be neglected and hence dropping packets by no-buffer policy is also near-optimal.

In Fig. \ref{fig_ipra_perf}, we simulate IPRA, along with the centralized index policy given by Theorem \ref{thm2} which has been shown near-optimal according to Fig. \ref{fig_mdp} and \ref{fig_indN}. In IPRA, the index threshold and transmission probability should be optimized to achieve good performance. Herein, however, we do not bother to choose the optimal parameter and use a very simple approach: we fix the transmission probability to be $0.2$ and optimize the index threshold based on a one-dimensional search. In Fig. \ref{fig_ipra_perf} (b), we observe that the AoI of IPRA (normalized by the AoI achieved by the centralized policy) approaches one as the packet transmission time increases. This indicates that IPRA is asymptotically optimal when the packet transmission time is large. The AoI fluctuates when $N$ is small because IPRA relies on that there are always multiple terminals with AoI larger than the threshold, which can fluctuate more with a smaller number of terminals. As $N$ increases, the performance of IPRA gets better since there are almost certainly terminals whose AoI is larger than the threshold.  Although it is found that in Fig. \ref{fig_ipra_perf} (a), the absolute AoI degradation is larger with a larger number of terminals, the normalized AoI asymptotic optimality is guaranteed (see Fig. \ref{fig_ipra_perf} (b)). 

\section{Concluding Remarks}
\label{sec_con}
In this paper, AoI is optimized in a wireless-connected star-topology network for massive machine-type status update sources. We propose a contention-based status update scheme and simulation results indicate that the performance of the proposed scheme is asymptotically optimal. This is based on the following results: First, the centralized Whittle's index policy which has near-optimal performance is derived in closed-form and the indexability is established. The performance thereof serves as a performance benchmark for optimality. Generalizations to stochastic and periodic packet arrivals and unreliable channels are also provided. Secondly, IPRA adopts the derived index to prioritize packet transmissions among terminals and a random access procedure is applied with asymptotically optimal performance. It is shown through computer-based simulations that the performance of IPRA is close to the centralized index policy and outperforms standard CSMA/CA schemes. In scenarios where hard AoI deadlines are enforced, we provide closed-form achievable AoI stationary distribution which indicates the maximum number of terminals that the system can support, with any given deadlines and their violation probabilities. 

An important future direction is to consider random access protocols for AoI minimization in general wireless networks. Previously when the throughput-optimality was considered, the line of work, e.g., \cite{raj09,ni12}, leveraged a Glauber dynamics based argument to show that a CSMA-based protocol is throughput-optimal in general; however, this argument cannot be generalized to AoI minimization problems. Novel random access designs and proof techniques are thus required.

\appendices
\section{Proof for Theorem \ref{thm1}}
\label{app_thm1}
For ease of reference, we restate \cite[Proposition 4.6.1]{bertsekas2007dynamic} here, 
\begin{prop}
\label{prop4.6.1}
Let the state spaces $S$ be countably infinite. Assume that a scalar $\hat{J}^*$ and a real-valued function $f$ solve Bellman's equation, i.e., for all states $i$,
\begin{iarray}
\label{4.121}
    \hat{J}^* + f(i) = \min_{u\in U(i)} \left[ g(i,u) + \sum_{j\in S}p_{ij}(u)f(j)\right],
\end{iarray}
and furthermore $f$ satisfies for all policies $\pi$ and states $i$
\begin{iarray}
\label{4.122}
    \lim_{N\to\infty}\frac{1}{N} E\left[ f(x_N) | x_0 = i, \pi\right] = 0.
\end{iarray}
Then
\begin{iarray}
    \hat{J}^* = \min_\pi J_\pi.
\end{iarray}
Furthermore, if $\mu^{*}(i)$ attains the minimum in \eqref{4.121} for each $i$, the stationary policy $\mu^{*}$ is optimal.
$\hfill\square$
\end{prop}

In the following, we first assume that the solution of Bellman's equation and its associated policy satisfy two conditions. Then we can obtain $\hat{J}^*$ and $f$. Checking \eqref{4.121} and \eqref{4.122} concludes the proof.

As a first step, given an arbitrary auxiliary service charge $m \ge 0$, two conditions regarding the solution and its associated optimal policy are claimed.
\begin{condition}
\label{ass2}
Bellman's equation has solution $\hat{J}^*$ and function $f$ which satisfies $f(a,0) \le f(a,1) \le\cdots\le f(a,d) \le\cdots$ for any $a$,. $\hfill\square$
\end{condition}
\begin{condition}
\label{ass1}
$\hat{J}^*$ and function $f$ can be produced by a threshold policy $\boldsymbol{\pi}_{\mathsf{D}}$ which, in turn, attains the minimum in Bellman's equation. With policy $\boldsymbol{\pi}_{\mathsf{D}}$, the optimal action of state $(a,d)$ is to idle when $0 \le d < D_a$ and to schedule when $d \ge D_a$. Furthermore, the thresholds satisfy $D_1 \le D_2 \le\cdots\le D_a \le\cdots$. $\hfill\square$
\end{condition}


We first prove that thresholds ${D_a}$ converge.
\begin{lemma}
\label{lm1}
There exists $a_{\mathsf{M}}$ such that $\forall a \ge a_{\mathsf{M}}$, the threshold $D_a=D_{a_\mathsf{M}}$ is a constant. $\hfill\square$
\end{lemma}
\begin{IEEEproof}
Based on Condition \ref{ass1}, the thresholds are monotonically increasing with respect to $a$. Note that when $d \ge m$, observing the cost-to-go equations in \eqref{c2go}, the optimal action in this case is to schedule the terminal. Specifically, with $d \ge m$, 
\begin{iarray}
    && \mu_0(a,d) - \mu_1(a,d) \nonumber\\
    &=& d-m + (1-\lambda)(f(a+1,d)-f(a+1,0))  + \lambda(f(1,d+a)-f(1,a)) \ge 0.
\end{iarray}
The inequality is based on Condition \ref{ass2}. This means that $\forall a$, $D_a \le m$. Therefore, the threshold array $\{D_a:\,a=1,2,\cdots\}$ is monotonically increasing with a finite upper bound; hence the limitation exists. Since $D_a$ is integer, the limitation is attainable and this concludes the proof.
\end{IEEEproof}

With $\boldsymbol{\pi}_{\mathsf{D}}$, an important property regarding the differential cost-to-go function is stated below. 
\begin{prop}
\label{prop1}
For any $a_1,a_2\ge1$, $0 \le d_1< D_{a_1}$, $0 \le d_2< D_{a_2}$, and $a_1+d_1=a_2+d_2$, we have $f(a_1,d_1)=f(a_2,d_2)$. $\hfill\square$
\end{prop}
\begin{IEEEproof}
Given $\boldsymbol{\pi}_{\mathsf{D}}$ and based on Lemma \ref{lm1}, the action when $a \ge a_{\mathsf{M}}$ and $d \ge D_{a_\mathsf{M}}$ is to schedule the terminal. Therefore, based on \eqref{c2go} and Condition \ref{ass1} we have $D_{a_\mathsf{M}} \ge D_1$, and hence
\begin{equation}
\label{1d}
    f(1,d)+ \hat{J}^* = 1+m+(1 - \lambda )f(2,0) + \lambda f(1,1),\, d \ge D_{a_\mathsf{M}}.
\end{equation}
Note that based on \eqref{1d}, $f(1,d) = f(1,D_{a_\mathsf{M}})$, $\forall d \ge D_{a_\mathsf{M}}$. Additionally, the action when $d = D_{a_\mathsf{M}}-1$ and $a \ge a_{\mathsf{M}}$ is to idle, and hence,
\begin{iarray}
     f(a,D_{a_\mathsf{M}}-1) + \hat{J}^* &=& D_{a_\mathsf{M}}+a-1+ (1 - \lambda )f(a + 1,D_{a_\mathsf{M}}-1)  + \lambda f(1,D_{a_\mathsf{M}} + a-1) \nonumber\\
    &=& D_{a_\mathsf{M}}+a-1+ (1 - \lambda )f(a + 1,D_{a_\mathsf{M}}-1)  +  \lambda f(1,D_{a_\mathsf{M}}).
\end{iarray}
Denote $g \triangleq D_{a_\mathsf{M}}- \hat{J}^* -1 + \lambda f(1,D_{a_\mathsf{M}})$, we obtain
\begin{iarray}
\label{rec}
    && f(a,D_{a_\mathsf{M}}-1) = a + (1 - \lambda )f(a + 1,D_{a_\mathsf{M}}-1) +  g.
\end{iarray}
Solving \eqref{rec} recursively with respect to $a$ yields
\begin{iarray}
\label{drope}
    && f(a,D_{a_\mathsf{M}}-1) = \gamma_0{(1 - \lambda )^{ a_{\mathsf{M}} - a}} + \frac{a+g}{\lambda}+\frac{1-\lambda}{\lambda^2}
\end{iarray}
where $\gamma_0 = f(a_{\mathsf{M}},D_{a_\mathsf{M}}-1) - \frac{a_\mathsf{M}+g}{{\lambda}} - \frac{1-\lambda}{\lambda^2}$ and $a \ge a_{\mathsf{M}}$. 
We will show the solution with $\gamma_0=0$ is a valid solution by checking the consistency. Likewise, we obtain
\begin{iarray}
\label{ds}
    &&f(a,D_{a_\mathsf{M}}-s) = \frac{a+g-s}{\lambda}+\frac{1}{\lambda^2}, \,a \ge s+a_{\mathsf{M}}-1,\, 1 \le s \le D_{a_\mathsf{M}}.
\end{iarray}
Note that, e.g., $f(a_{\mathsf{M}},D_{a_\mathsf{M}}-1) = f(a_{\mathsf{M}}+1,D_{a_\mathsf{M}}-2) = \cdots= f(D_{a_\mathsf{M}}+a_{\mathsf{M}}-1,0)$. It is therefore clear that the proposition holds for any $a_1,a_2,d_1,d_2$ satisfying $a_1+d_1=a_2+d_2 \ge D_{a_\mathsf{M}}+a_{\mathsf{M}}-1$. For $a_1,a_2,d_1,d_2$ satisfying $a_1+d_1=a_2+d_2 < D_{a_\mathsf{M}}+a_{\mathsf{M}}-1$, an induction based proof is adopted. Suppose the proposition holds for any $a_1,a_2,d_1,d_2$ satisfying $a_1+d_1=a_2+d_2 = D_{a_\mathsf{M}}+a_{\mathsf{M}}-1-s$, $0 \le s \le D_{a_\mathsf{M}}+a_{\mathsf{M}}-2$, $d_1< D_{a_1}$ and $d_2< D_{a_2}$, then for any $a_1^\prime,a_2^\prime,d_1^\prime,d_2^\prime$ satisfying $a_1^\prime+d_1^\prime=a_2^\prime+d_2^\prime = D_{a_\mathsf{M}}+a_{\mathsf{M}}-2-s$ and $d_1^\prime< D_{a_1^\prime}$, $d_2^\prime< D_{a_2^\prime}$, the action is to idle based on $\boldsymbol{\pi}_{\mathsf{D}}$. It follows from \eqref{c2go} that
\begin{iarray}
    f(a_1^\prime, d_1^\prime) &=& -\hat{J}^*+d_1^\prime + a_1^\prime + (1 - \lambda )f(a_1^\prime + 1,d_1^\prime) + \lambda f(1,d_1^\prime + a_1^\prime) \nonumber\\
    & \overset{(a)}{=} & -\hat{J}^*+d_2^\prime + a_2^\prime + (1 - \lambda )f(a_2^\prime + 1,d_2^\prime) + \lambda f(1,d_2^\prime + a_2^\prime) = f(a_2^\prime, d_2^\prime).
\end{iarray}
The equality $(a)$ is based on the induction hypothesis. Also note that $d_1^\prime< D_{a_1^\prime} \le D_{a_1^\prime+1}$ based on Condition \ref{ass1}, and hence the conditions are all satisfied. For the induction basis, the proposition holds for $s=0$ based on \eqref{ds}. Therefore, the proposition is concluded.
\end{IEEEproof}

To proceed, we obtain another important property of the solution in the following proposition. 
\begin{prop}
\label{prop2}
For any state $(a,d)$ with $a \ge 1$ and $d \ge D_a$, $f(a,d) - f(a,0) = m$. $\hfill\square$
\end{prop}
\begin{IEEEproof}
For some state $(a,d)$ with $a \ge 1$ and $d \ge D_a$, based on \eqref{c2go} and $\boldsymbol{\pi}_{\mathsf{D}}$,
\begin{iarray}
f(a,d) &=& -\hat{J}^* + a +m + (1-\lambda)f(a+1,0) + \lambda f(1,a) \nonumber\\
f(a,0) &=& -\hat{J}^* + a + (1-\lambda)f(a+1,0) + \lambda f(1,a).
\end{iarray}
Therefore the proposition is concluded by observing the difference of the two above equations.
\end{IEEEproof}

Resuming the proof of the theorem, for $1 \le a < D_1$, based on \eqref{c2go},
\begin{iarray}
    f(a,0) &=& -\hat{J}^*+a + (1 - \lambda ) f(a+1,0) + \lambda f(1,a) = -\hat{J}^*+a + f(a+1,0),
\end{iarray}
where the last equality is based on Proposition \ref{prop1}. Given that $f(1,0)=0$, it follows that
\begin{equation}
\label{ds3}
    f(a,0) = (a-1)\hat{J}^* - \frac{a(a-1)}{2}, \, 1 \le a < D_1+1.
\end{equation}

Based on Proposition \ref{prop1} and similar arguments in \eqref{drope}, we obtain
\begin{equation}
\label{ds2}
    f(a,0) = \frac{a-\hat{J}^*-1}{\lambda}+m+\frac{1}{\lambda^2}, \, a \ge D_1.
\end{equation}
Based on Proposition \ref{prop2}, when $d \ge D_a$ it follows that
\begin{iarray}
\label{2d}
     f(a,d) &=& m+ f(a,0) = \left\{\,
        \begin{IEEEeqnarraybox}[][c]{l?s}
        \IEEEstrut	
        m + a \hat{J}^* - \frac{(a-1)a}{2}, &  if $1 \le a < D_1+1$;\\
    	\frac{a}{\lambda}+2m-\frac{1-\lambda}{\lambda}\hat{J}^* + \frac{1-\lambda}{\lambda^2}, & if $a \ge D_1$.
    	\IEEEstrut
    	\end{IEEEeqnarraybox}
    	\right.    	
\end{iarray}
Combining \eqref{ds3} and \eqref{ds2} when $d=D_1$ gives us the relationship among $m$, $\hat{J}^*$ and $D_1$, i.e.,
\begin{equation}
    \label{mjd}
    m = \left(D_1-1+\frac{1}{\lambda}\right) \hat{J}^* -\frac{D_1^2}{2} + \frac{D_1}{2} -\frac{D_1}{\lambda} + \frac{\lambda-1}{\lambda^2}.
\end{equation}

Now we have obtained all differential cost-to-go function expressions based on $\boldsymbol{\pi}_{\mathsf{D}}$. To obtain $\hat{J}^*$ and $\{D_a\}$, we resort to Condition \ref{ass1} that $\boldsymbol{\pi}_{\mathsf{D}}$ attains the minimum in Bellman's equation.

For state $(a,D_a)$, the action under policy $\boldsymbol{\pi}_{\mathsf{D}}$ is to schedule. Since $\boldsymbol{\pi}_{\mathsf{D}}$ attains the minimum, the upper part in the minimization (denoted by $\mu_0(a,D_a)$) is larger than the lower part (denoted by $\mu_1(a,D_a)$). First, consider the case where $a \ge D_1$, the difference is
\begin{iarray}
\label{48}
    && \mu_0(a,D_a) - \mu_1(a,D_a) \nonumber\\
    &=& D_a-m + (1-\lambda)(f(a+1,D_a)-f(a+1,0))  + \lambda(f(1,D_a+a)-f(1,a)) \nonumber\\
    &\overset{(a)}{=}& D_a-m + (1-\lambda)\frac{D_a}{
    \lambda} \ge0,
\end{iarray}
where the equality $(a)$ is attributed to \eqref{1d} and \eqref{ds2}. Therefore we obtain a condition $D_a \ge \lambda m$. The case with $1\le a < D_1$ yields
\begin{iarray}
\label{49}
    && \mu_0(a,D_a) - \mu_1(a,D_a) \nonumber\\
    &=& D_a-m + (1-\lambda)(f(a+1,D_a)-f(a+1,0))  + \lambda(f(1,D_a+a)-f(1,a)) \nonumber\\
     &\overset{(a)}{=}& D_a-m + (1-\lambda)f(a+1,D_a) + \lambda m -f(a+1,0)\nonumber\\
    &\overset{(b)}{=}& D_a-m + (1-\lambda)\left(\frac{D_a+a-\hat{J}^*}{\lambda} +m + \frac{1}{\lambda^2}\right)  + \lambda m - a\hat{J}^* + \frac{a(a+1)}{2} \ge 0,
\end{iarray}
where the equality $(a)$ follows from Proposition \ref{prop1} and \ref{prop2} and equality $(b)$ is from \eqref{ds2}. We obtain the second condition $D_a \ge   (1-\lambda+a\lambda)\hat{J}^* - a + 1 - \lambda \frac{a(a-1)}{2} -\frac{1}{\lambda}$. 

For states $(a,D_a-1)$ with $a \ge D_1$, the terminal should be idle based on $\boldsymbol{\pi}_{\mathsf{D}}$. We obtain
\begin{iarray}
\label{50}
    && \mu_0(a,D_a-1) - \mu_1(a,D_a-1) \nonumber\\
    &=& D_a-1-m + (1-\lambda)(f(a+1,D_a-1)-f(a+1,0))  + \lambda(f(1,D_a-1+a)-f(1,a)) \nonumber\\
    &\overset{(a)}{=}& D_a-1-m + \frac{1-\lambda}{\lambda}(D_a-1) \le 0,
\end{iarray}
where the equality $(a)$ follows from \eqref{ds2}, and thus another condition is $D_a \le \lambda m + 1$. With $1 \le a < D_1$,
\begin{iarray}
\label{51}
    && \mu_0(a,D_a-1) - \mu_1(a,D_a-1) \nonumber\\
    &=& D_a-1-m + (1-\lambda)(f(a+1,D_a-1)-f(a+1,0))  + \lambda(f(1,D_a-1+a)-f(1,a)) \nonumber\\
    &\overset{(a)}{=}& D_a-1-m + (1-\lambda)f(a+1,D_a-1)  + \lambda m -f(a+1,0)\nonumber\\
    &\overset{(b)}{=}& D_a-1-m + (1-\lambda)\left(\frac{D_a+a-\hat{J}^*-1}{\lambda} +m + \frac{1}{\lambda^2}\right)  + \lambda m - a\hat{J}^* + \frac{a(a+1)}{2} \le 0,
\end{iarray}
where the equality $(a)$ follows from Proposition \ref{prop1} and \ref{prop2} and equality $(b)$ is from \eqref{ds2}. Hence, the following condition should be satisfied $D_a \le   (1-\lambda+a\lambda)\hat{J}^* - a +2 - \lambda \frac{a(a-1)}{2} -\frac{1}{\lambda}$. 
Since the thresholds are integer, they are summarized as following 
\begin{iarray}
\label{thres}
     D_a = \left\{\,
        \begin{IEEEeqnarraybox}[][c]{l?s}
        \IEEEstrut	
        \left\lceil (1-\lambda+a\lambda)\hat{J}^* - a + 1 - \lambda \frac{a(a-1)}{2} -\frac{1}{\lambda} \right\rceil, &  if $1 \le a < D_1$;\\
    	\left\lceil \lambda m \right\rceil, & if $a \ge D_1$.
    	\IEEEstrut
    	\end{IEEEeqnarraybox}
    	\right.    	
\end{iarray}
We obtain $D_1 = \left\lceil \hat{J}^* -\frac{1}{\lambda} \right\rceil$ from  \eqref{thres} since all thresholds must be larger than 1 at least. Combining this and \eqref{mjd}, the average AoI $\hat{J}^*$ can be computed by solving the equation. 

Thus far, we have obtained $\hat{J}^*$, $f(a,d)$ and policy $\boldsymbol{\pi}_{\mathsf{D}}$ based on Condition \ref{ass2} and \ref{ass1}. It is straightforward to check that $\hat{J}^*$, $f(a,d)$ is the solution to \eqref{c2go} and $\boldsymbol{\pi}_{\mathsf{D}}$ attains the minimum in the right hand side of $\boldsymbol{\pi}_{\mathsf{D}}$. What is still left to do is to prove that $f(a,d)$ satisfies \eqref{4.122} in Proposition \ref{prop4.6.1}.

The packet age $a$ forms a Markov chain which is independent of the policy $\boldsymbol{\pi}$. The stationary distribution of $a$ is geometric distribution with success probability $\lambda$, that is $P(a = i) = (1-\lambda)^{i-1}\lambda$. With the expression of $f(a,d)$, we can verify that $f(a,d) \le f(a,D_{a_M})$. Thus, \eqref{4.122} can be proved as
\begin{iarray}
    &&\lim_{N\to\infty}\frac{1}{N} \mathbb{E}\left[ f(a(N),d(N)) | a(0),d(0), \pi\right] \nonumber \\
    &\le& \lim_{N\to\infty}\frac{1}{N} \mathbb{E}\left[ f(a(N),D_{a_M}) | a(0),d(0), \pi\right] \nonumber\\
    &=&  \lim_{N\to\infty}\frac{1}{N} \mathbb{E}\left[ f(a(N),D_{a_M}) | a(0)\right] \nonumber \\
    &=& \lim_{N\to\infty}\frac{1}{N} \mathbb{E}\left[ f(a,D_{a_M})\right] \nonumber \\
    &=& 0.
\end{iarray}
Hence, we confirm that $\hat{J}^*$, $f(a,d)$ is solution of the Bellman's equation and the threshold policy $\boldsymbol{\pi}_{\mathsf{D}}$ is optimal, which concludes the proof.

\section{Proof for Theorem \ref{thm2}}
\label{app_thm2}
Observe the thresholds in Theorem \ref{thm1}. With $m=0$, the maximum threshold equals zero. Moreover, based on the monotonicity of the thresholds which is shown in the proof of Theorem \ref{thm1}, all thresholds are zero and hence the idle state space is empty. On the other hand, when $m$ goes to infinity, all the thresholds go to infinity with it and hence the idle state space approaches the entire space. Moreover, for any $m_1<m_2$, the thresholds associated with $m_1$ are no larger than those with $m_2$, and hence the monotonicity condition follows straightforwardly. This concludes the proof.

\section{Proof of Corollary \ref{coro_e}}
\label{app_coro_e}
We use the index with Bernoulli arrivals to demonstrate the proof technique. The case with periodic arrivals can be derived similarly. The cost-to-go function is modified from \eqref{c2go} to
\begin{iarray}
\label{c2g2}
 f(a,d) + \hat{J}^*  &=&  \min \left\{d + a + (1 - \lambda )f(a + 1,d) + \lambda f(1,d + a),\right. \nonumber\\ 
&& m+ (1-p_\mathsf{e}) \left(a + (1 - \lambda )f(a + 1,0) + \lambda f(1,a)\right) \nonumber\\
&& \left. + p_\mathsf{e} \left(d+a + (1 - \lambda )f(a + 1,d) + \lambda f(1,d + a)\right) \right\}, 
\end{iarray}
Denote the first entry in the minimization of \eqref{c2g2} as $\mu_{\mathsf{e},0}(a,d)$ and the second $\mu_{\mathsf{e},1}(a,d)$. Following the same procedure as in the proof of Theorem \ref{thm1}, we obtain that when $d=D_a$, both decisions are equally beneficial, namely $\mu_{\mathsf{e},0}$ should be equal to $\mu_{\mathsf{e},1}$; this can be observed from \eqref{48}-\eqref{51}. Hence,
\begin{iarray}
\label{efd}
    && \mu_{\mathsf{e},0}(a,D_a) = \mu_{\mathsf{e},1}(a,D_a) \Longleftrightarrow \nonumber\\
    && D_a  + (1 - \lambda )(f(a + 1,D_a)-f(a + 1,0)) + \lambda (f(1,D_a + a)-f(1,a))= \frac{m}{1-p_\mathsf{e}}. 
\end{iarray}
Moreover, when $d=D_a$, the decision is to schedule; when $d=0$, the decision is to idle. The cost-to-go function yields
\begin{iarray}
\label{mp}
&& f(a,D_a) - f(a,0) \nonumber\\
&=& m + p_\mathsf{e} D_a +p_\mathsf{e}\left((1 - \lambda )(f(a + 1,D_a)-f(a + 1,0)) + \lambda (f(1,D_a + a)-f(1,a))\right) \nonumber\\
&\overset{(a)}{=}& m+ p_\mathsf{e} \frac{m}{1-p_\mathsf{e}} = \frac{m}{1-p_\mathsf{e}}.
\end{iarray}
The equality $(a)$ follows from \eqref{efd}. For $d>D_a$, the decision is to schedule, therefore we obtain
\begin{iarray}
\label{c1}
&& f(a,d+1) - f(a,d) \nonumber\\
&=& p_\mathsf{e}(1 + (1 - \lambda )(f(a + 1,d+1)-f(a + 1,d)) + \lambda (f(1,a+d+1)-f(1,a+d)) ).
\end{iarray}
When $p_\mathsf{e}$ is small, it follows that 
\begin{equation}
\label{c2}
f(a,d+1) - f(a,d) = \smallO{(1)}.
\end{equation}
Therefore, combining with \eqref{mp} we have $f(a,d) = \frac{m}{1-p_\mathsf{e}} + \smallO{(1)}\textrm{, when }d \ge D_a\textrm{ and }p_\mathsf{e} \to 0$. With this, we can obtain the Whittle's index with unreliable channels following the same steps as in Appendix \ref{app_thm1}. 
The indexability also follows because $\frac{m}{1-p_{\mathsf{e}}}$, compared with $m$, does not affect the monotonicity or index values at zero and infinity. This concludes the proof.

\section{Proof of Corollary \ref{coro_dl}}
\label{app_coro_dl}
Since the reliability requirement $\epsilon$ is arbitrarily small, we only need to consider the case when the scheduling interval $\Gamma$ is smaller than the AoI deadline.
Following from \eqref{eq_dist}, it follows that
\begin{iarray}
\label{le1}
    \Pr\{h_n \ge H_n\} \le \epsilon_n &\Longleftrightarrow& (1-\lambda_n)^{H_n-\Gamma_n+1} \frac{1-(1-\lambda_n)^{\Gamma_n}}{\lambda_n\Gamma_n} \le \epsilon_n \nonumber\\ 
    &\Longleftrightarrow& \frac{1-(1-\lambda_n)^{\Gamma_n}}{(1-\lambda_n)^{\Gamma_n} \Gamma_n} \le c(\lambda_n,H_n)\epsilon_n \Longleftrightarrow \Gamma_n \le v_n^{-1}(c\left(\lambda_n,H_n)\epsilon_n\right),
\end{iarray}
where $v_n(\Gamma_n)\triangleq\frac{1-(1-\lambda_n)^{\Gamma_n}}{(1-\lambda_n)^{\Gamma_n} \Gamma_n}$. It is straightforward to verify that $v_n(\Gamma_n)$ is monotonically non-decreasing in $[1,\infty)$ and hence its inverse function exists. In addition, given that $(1-\lambda_n)^{\Gamma_n} \ll 1$ in the high-reliability regime, it follows that $v_n(\Gamma_n) \approx \frac{1}{(1-\lambda_n)^{\Gamma_n} \Gamma_n}$ and therefore 
\begin{equation}
\label{last}
    \Gamma_n \le \frac{W_{-1}\left(\frac{\log(1-\lambda_n)}{\epsilon_n c(\lambda_n,H_n)}\right)}{\log(1-\lambda_n)}
\end{equation} by the definition of Lambert $W$ function. Therefore, the achievable reliable deadline region follows by plugging \eqref{le1} into $\sum_{n=1}^N \frac{1}{\Gamma_n} \le 1$.

Since the AoI scales with $N$, in the massive IoT regime where $N$ is large, it is reasonable that the deadline $H$ should also be large. Therefore, we adopt the following asymptotic results of the Lambert $W$ function $\lim_{x\to0^-}\frac{W_{-1}(x)}{\log (-x)}=1$. It follows from \eqref{last} that $\Gamma_n \le H -\frac{\log\epsilon}{\log(1-\lambda)} + c_0(\lambda)$, where $c_0(\lambda)\triangleq1+\frac{\log(-\log(1-\lambda))-\log \lambda}{\log(1-\lambda)}$. This inequality indicates that in the massive IoT regime, to achieve an AoI deadline of $H$, the total number of terminals $N_\mathsf{deadline}$ should satisfies \eqref{dl_mean}. This is proved by plugging the upper bound of $\Gamma_n$ in the necessary condition $\sum_{n=1}^{N} \frac{1}{\Gamma_n} \le 1$. The results for time-average AoI follows directly from the scaling results presented in \cite[Theorem 2]{jiang18_iot}. 

\bibliography{Ref.bib}
\bibliographystyle{IEEEtran}
\end{document}